\documentclass[fleqn,usenatbib]{mnras}
\usepackage[T1]{fontenc}
\usepackage{ae,aecompl}
\usepackage{graphicx}
\usepackage{amsmath}    
\usepackage{amssymb}    
\usepackage{float}

\usepackage{longtable}
\usepackage{pdflscape}

\title[Geometry transition in AGN jets]{A transition from parabolic to conical shape as a common effect in nearby AGN jets} 

\author[Kovalev et al.]{\parbox{\textwidth}{
Y.~Y.~Kovalev$^{1,2,3}$\thanks{E-mail: yyk@asc.rssi.ru},
A.~B.~Pushkarev$^{4,1}$,
E.~E.~Nokhrina$^{2}$,
A.~V.~Plavin$^{1,2}$,
V.~S.~Beskin$^{1,2}$,
A.V.~Chernoglazov$^{1,2}$,
M.~L.~Lister$^{5}$,
T.~Savolainen$^{6,7,3}$
}
\vspace{0.4cm}\\
\parbox{\textwidth}{
$^1$Lebedev Physical Institute, Leninsky prosp.~53, Moscow, 119991, Russia\\
$^2$Moscow Institute of Physics and Technology, Dolgoprudny, Institutsky per., 9, Moscow region, 141700, Russia\\
$^3$Max-Planck-Institut f\"ur Radioastronomie, Auf dem H\"ugel 69, 53121 Bonn, Germany\\
$^4$Crimean Astrophysical Observatory, Nauchny 298688, Crimea, Russia\\
$^5$Department of Physics and Astronomy, Purdue University, 525 Northwestern Avenue, West Lafayette, IN 47907, USA\\
$^6$Aalto University Mets\"ahovi Radio Observatory, Mets\"ahovintie 114, 02540 Kylm\"al\"a, Finland\\
$^7$Aalto University Department of Electronics and Nanoengineering, PL 15500, 00076, Aalto, Finland
}
}

\date{Accepted 2020 April 18. Received 2020 April 18; in original form 2019 June 23}

\pubyear{2020}

\begin{document}
\label{firstpage}
\pagerange{\pageref{firstpage}--\pageref{lastpage}}
\maketitle

\begin{abstract}
Observational studies of collimation in jets in active galactic nuclei (AGN) are a key to understanding their formation and acceleration processes.
We have performed an automated search for jet shape transitions in a sample of 367 AGN using VLBA data at 15~GHz and 1.4~GHz. 
This search has found ten out of 29 nearby jets at redshifts $z<0.07$ with a transition from a parabolic to conical shape, while the full analyzed sample is dominated by distant AGN with a typical $z\approx1$.
The ten AGN are 
UGC~00773, NGC~1052, 3C~111, 3C~120, TXS~0815$-$094, Mrk~180,  PKS~1514+00, NGC~6251, 3C~371, and BL~Lac. 
We conclude that the geometry transition may be a common effect in AGN jets. It can be observed only when sufficient linear resolution is obtained.
Supplementing these results with previously reported shape breaks in the nearby AGN 1H~0323+342 and M87, we estimate that the break occurs at $10^5$--$10^6$ gravitational radii from the nucleus.
We suggest that the jet shape transition happens when the bulk plasma kinetic energy flux becomes equal to the Poynting energy flux, while the ambient medium pressure is assumed to be governed by Bondi accretion.
In general, the break point may not coincide with the Bondi radius. The observational data supports our model predictions on the jet acceleration and properties of the break point.
\end{abstract}


\section{Introduction}
\label{s:intro}

Understanding the physical processes that determine the formation, acceleration and collimation of relativistic jets in active galactic nuclei (AGN) continues to be among the most challenging problems of modern astrophysics.
There are a wide variety of analytical and numerical models for jet acceleration and its confinement \citep[e.g.,][]{Vlahakis03,Beskin06,McKinney06,Komissarov07,Tchekhovskoy_11,McKinney12,PC15} that consider different solutions for jet shapes, such as cylindrical, conical and parabolic. 
General relativistic magnetohydrodynamic simulations
\citep[e.g.,][]{McKinney12}
predict that a jet starting from its apex has a parabolic streamline within the magnetically dominated acceleration zone. At other scales it transitions to a conical geometry associated with equipartition between energy densities of the magnetic field and the radiating particle populations. 
It has been shown for cold jets that acceleration should not occur in a conical jet. This requires something akin to  a parabolic jet shape closer to the jet base to allow differential expansion \citep{VK04,Komissarov_book12}.

In order to investigate these theories it is important to collect observational data on jet profile shapes for a large enough sample of AGN whose properties are well understood.
The first observational evidence for a transition from parabolic to conical jet shape was detected in M87 \citep{Asada12} at a distance of about 900\,mas near the feature HST-1, about 70~pc in projection, corresponding to $10^5$ Schwarzschild radii.
A few more studies of nearby AGNs to probe their innermost jet regions were performed recently: Mkn~501 \citep{Giroletti_etal08}, Centaurus~A \citep{r:CenA_TANAMI}, Cygnus~A \citep{Boccardi15_CygnusA,r:Nakahara19}, NGC\,6251 \citep{tseng16}, 1H~0323+342 \citep{Hada18}, 3C\,273 \citep{r:Akiyama18}, NGC\,4261 \citep{r:Nakahara18}, 3C\,84 \citep{RA_3C84}, 3C\,264 \citep{r:Boccardi_3C264}, NGC\,1052 \citep{r:Nakahara20}.
\citet{3C273_RM_ALMA} have indirectly addressed this question for the 3C\,273 jet close to the central engine on the basis of a model analysis of ALMA rotation measure data.
Larger survey studies \citep{Pushkarev09} have typically probed regions farther away from the central nucleus, although \citet{algaba17} have used apparent parsec-scale jet base parameters closer in.

\begin{figure*}
\vskip 4mm
\centering
\includegraphics[width=0.33\textwidth]{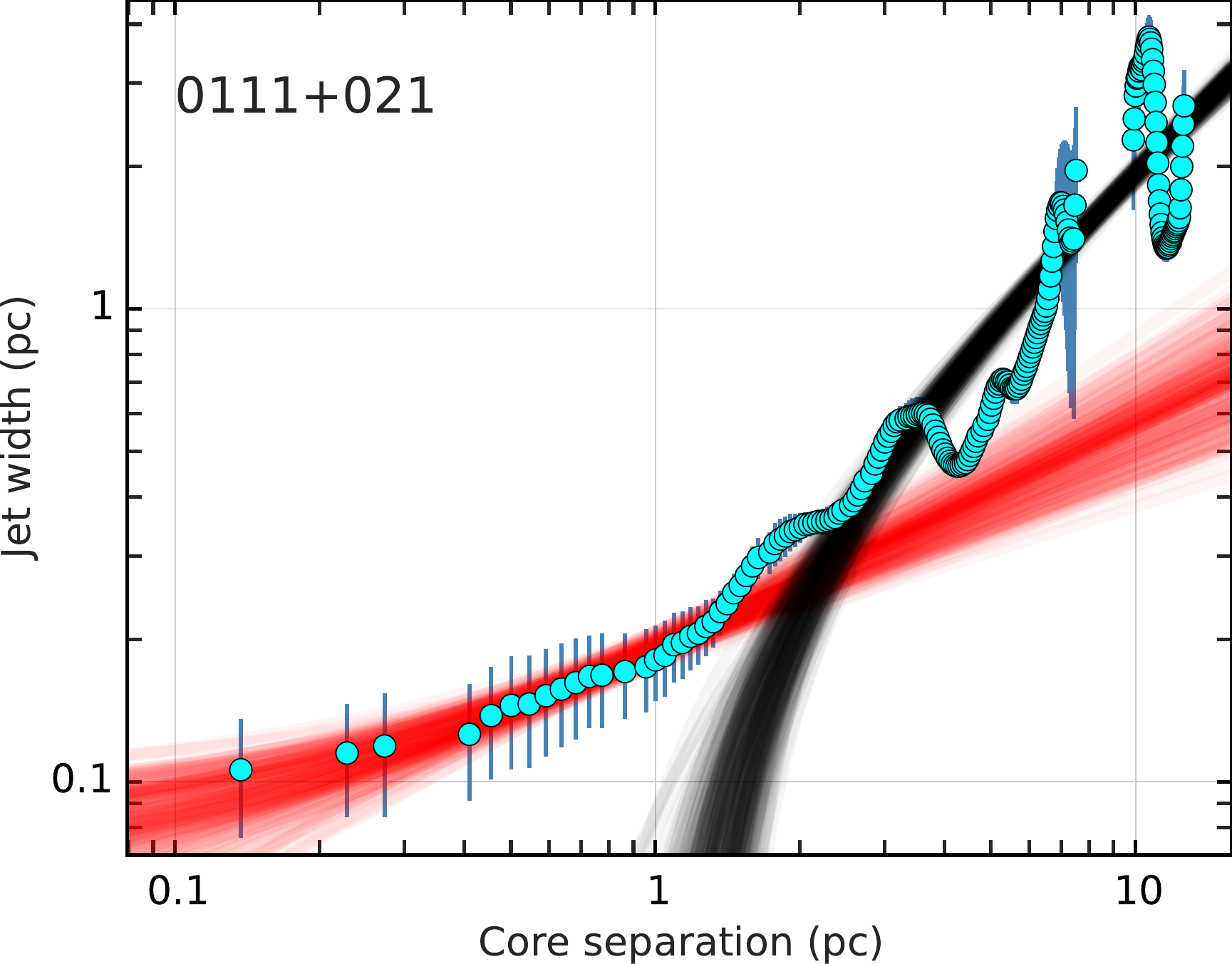}
\includegraphics[width=0.33\textwidth]{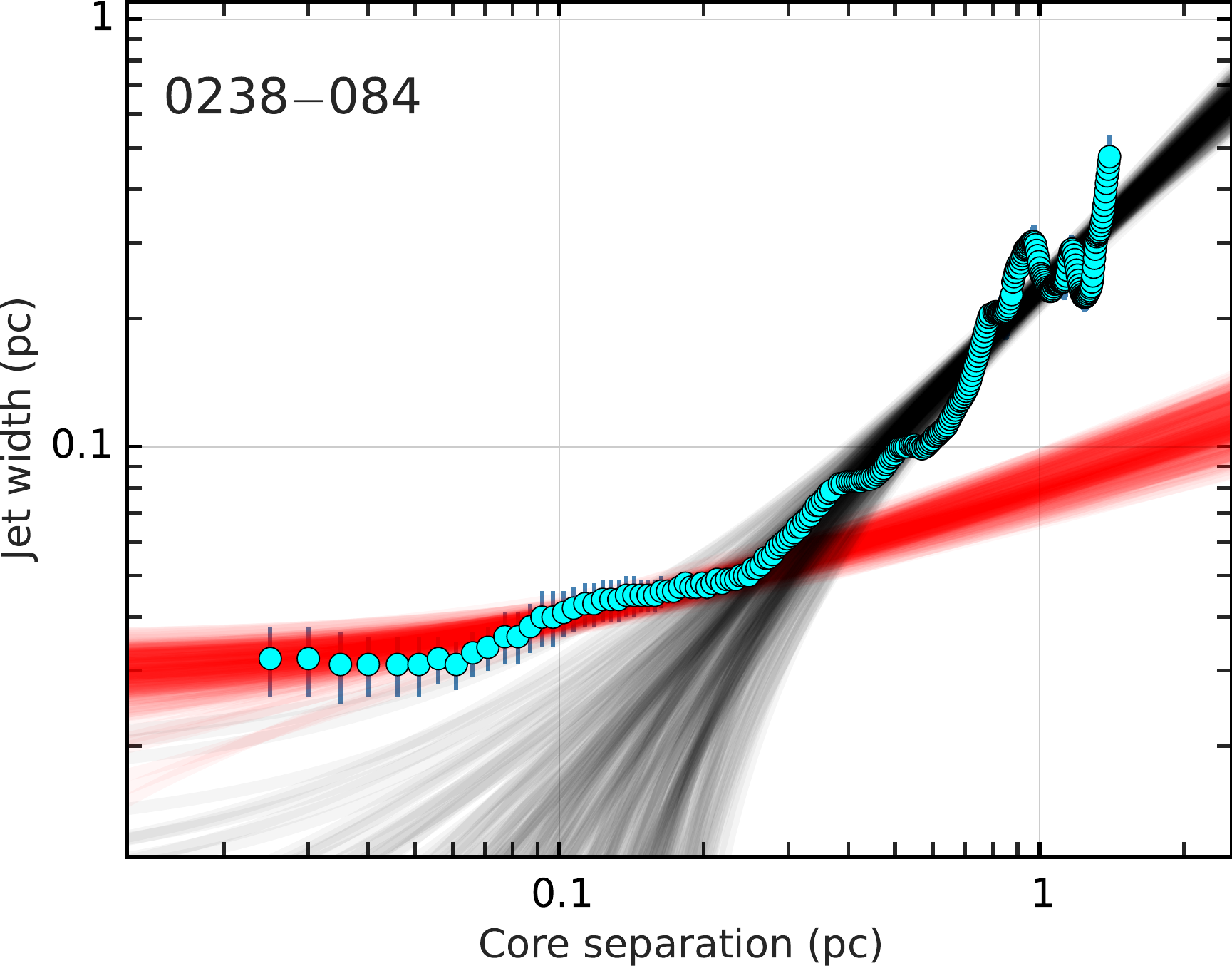}
\includegraphics[width=0.33\textwidth]{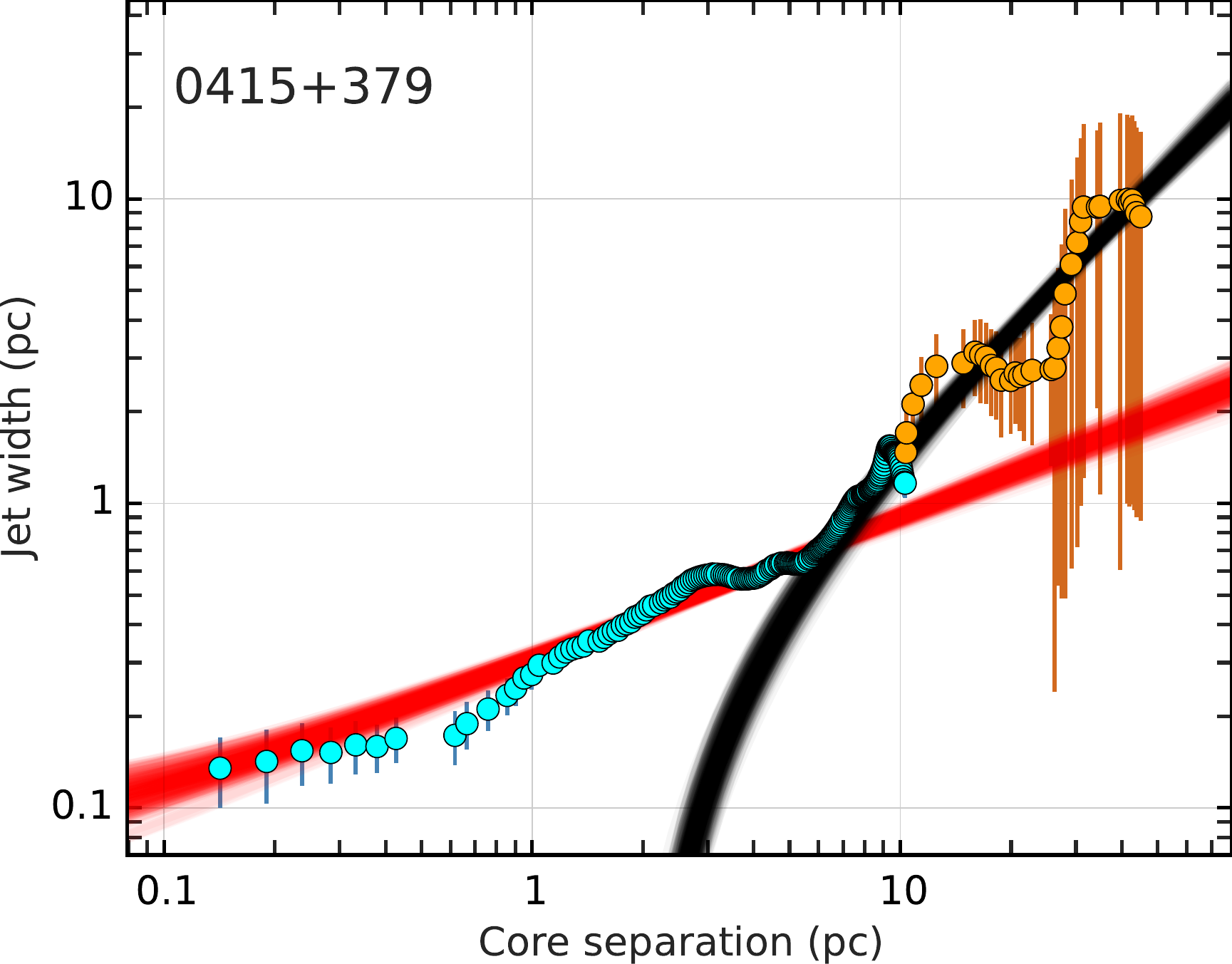}
\vskip 4mm
\includegraphics[width=0.33\textwidth]{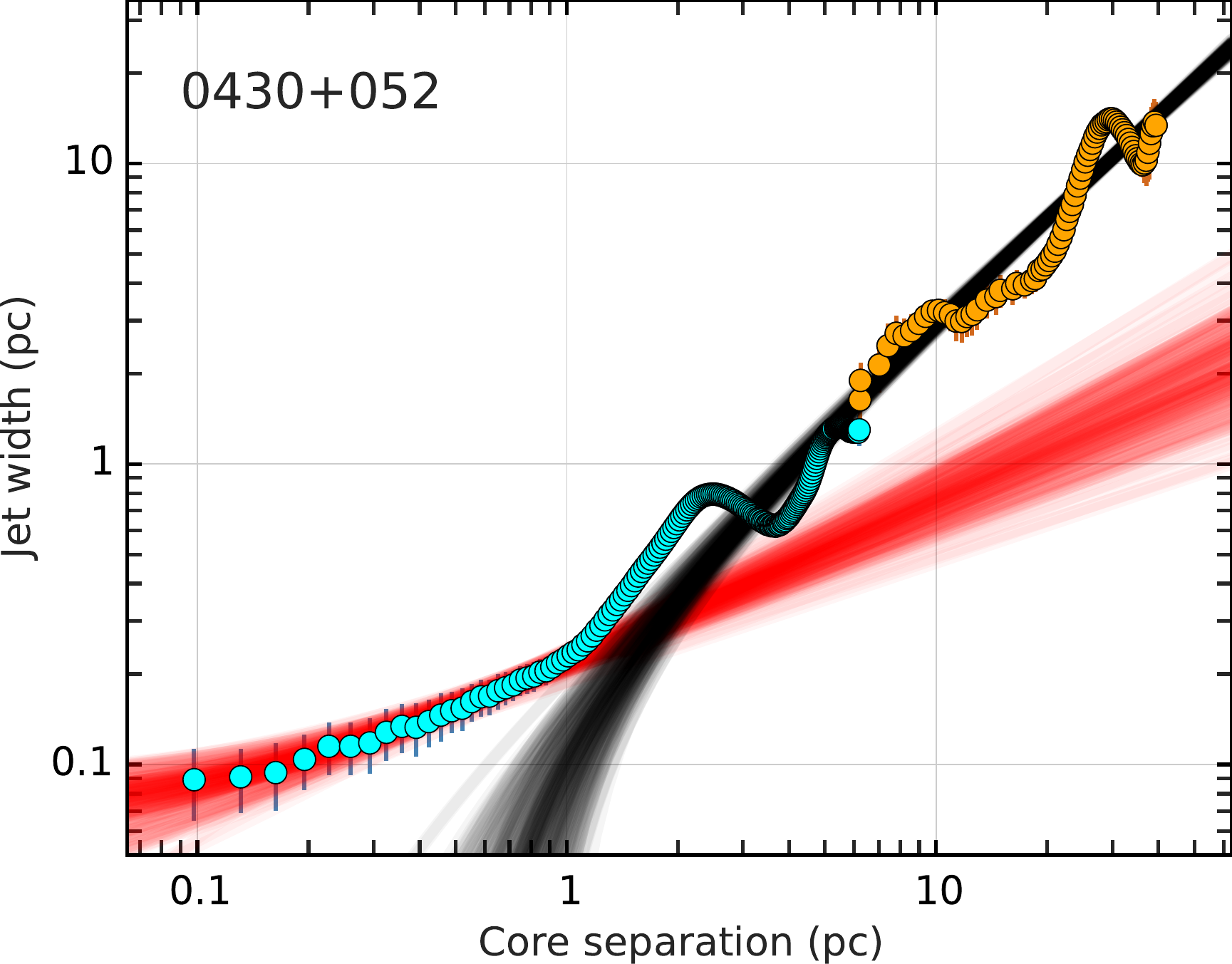}
\includegraphics[width=0.33\textwidth]{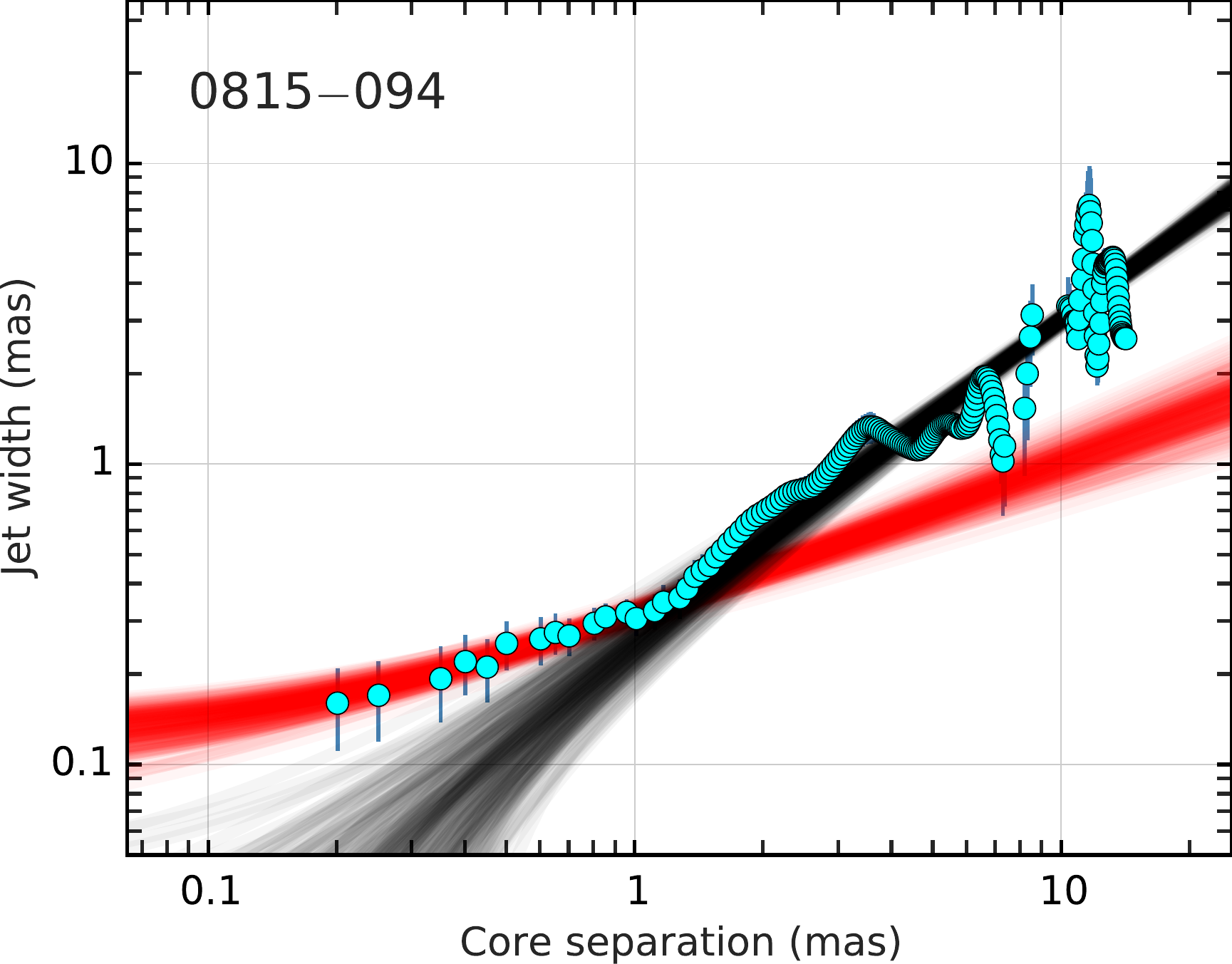}
\includegraphics[width=0.33\textwidth]{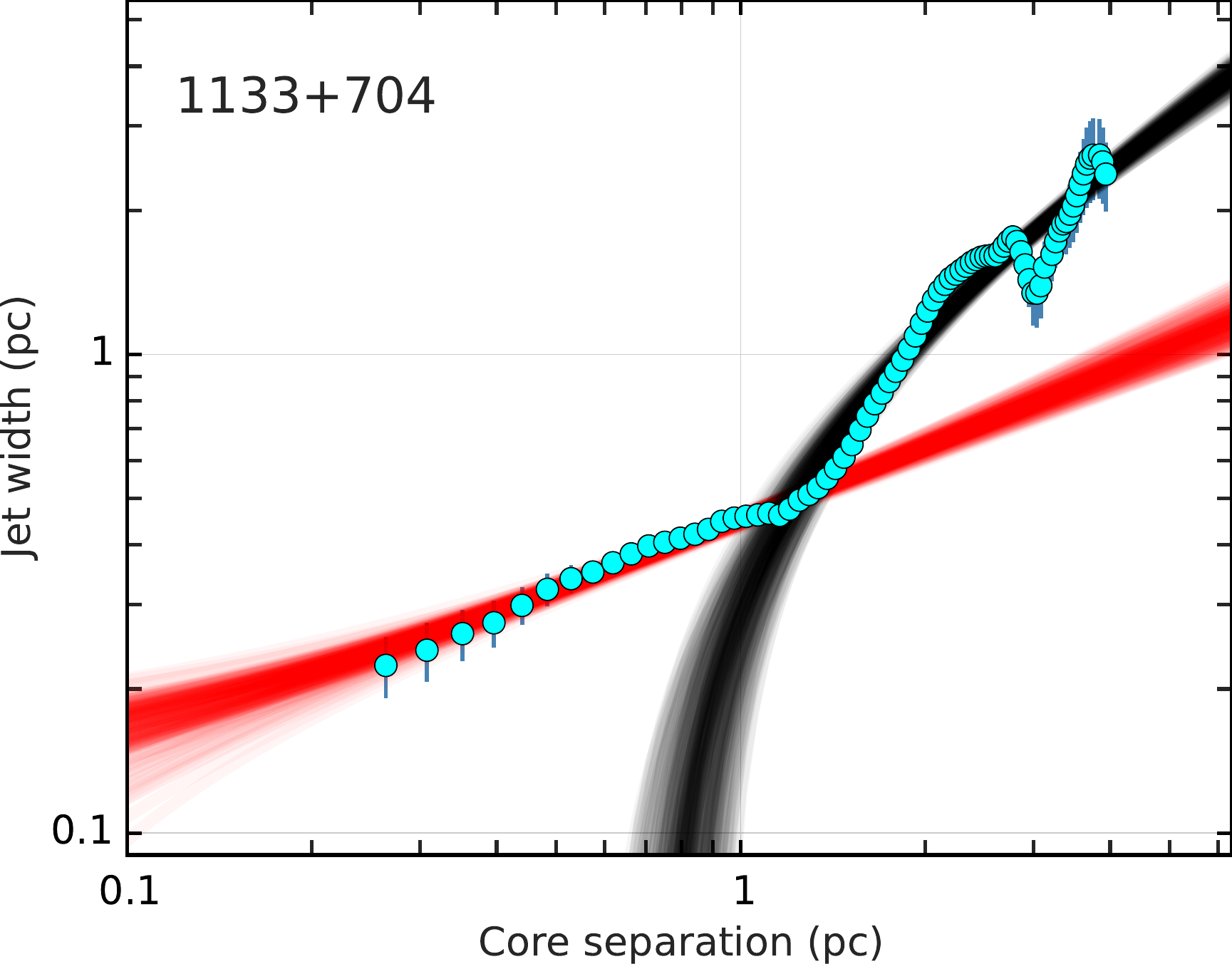}
\vskip 4mm
\includegraphics[width=0.33\textwidth]{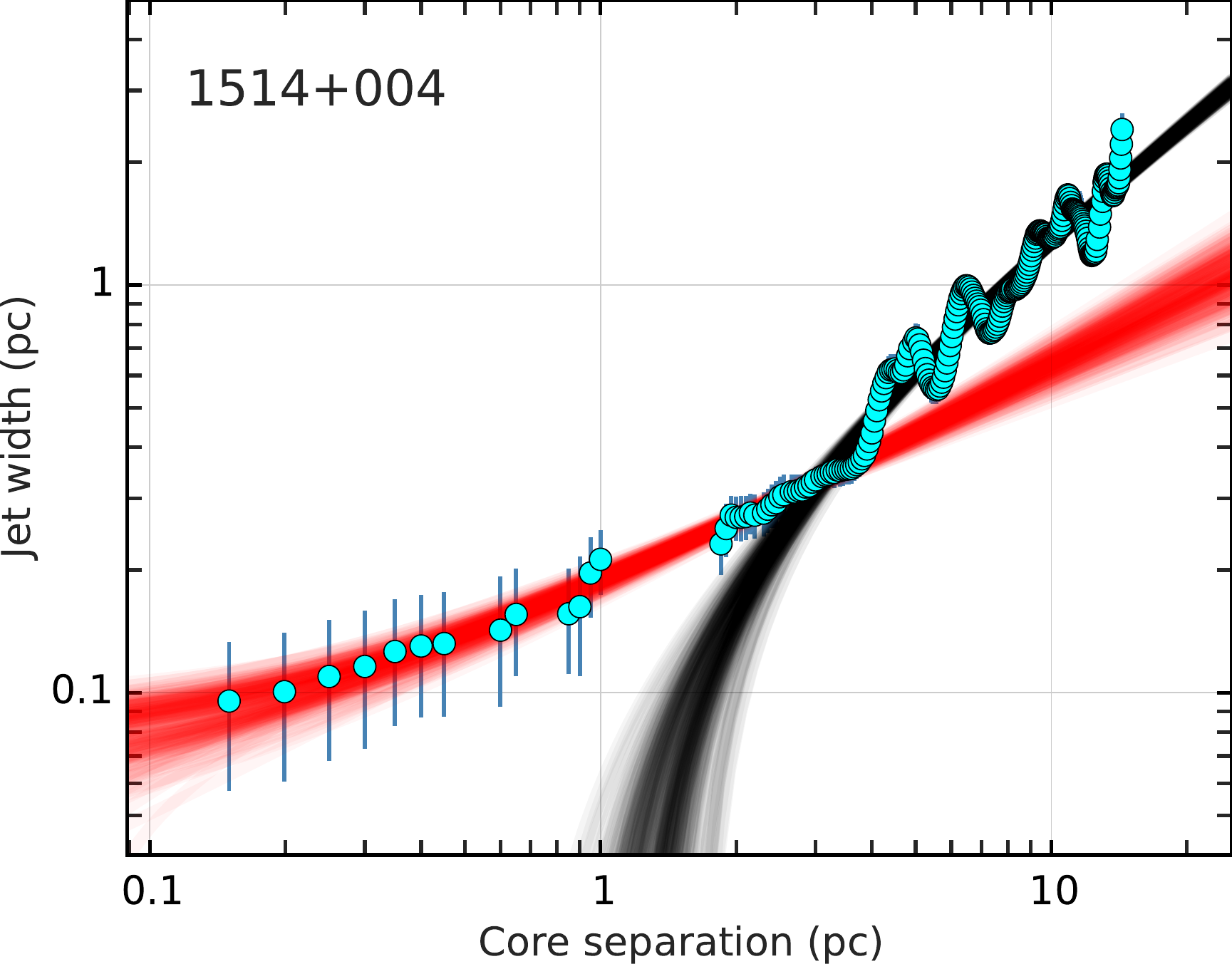}
\includegraphics[width=0.33\textwidth]{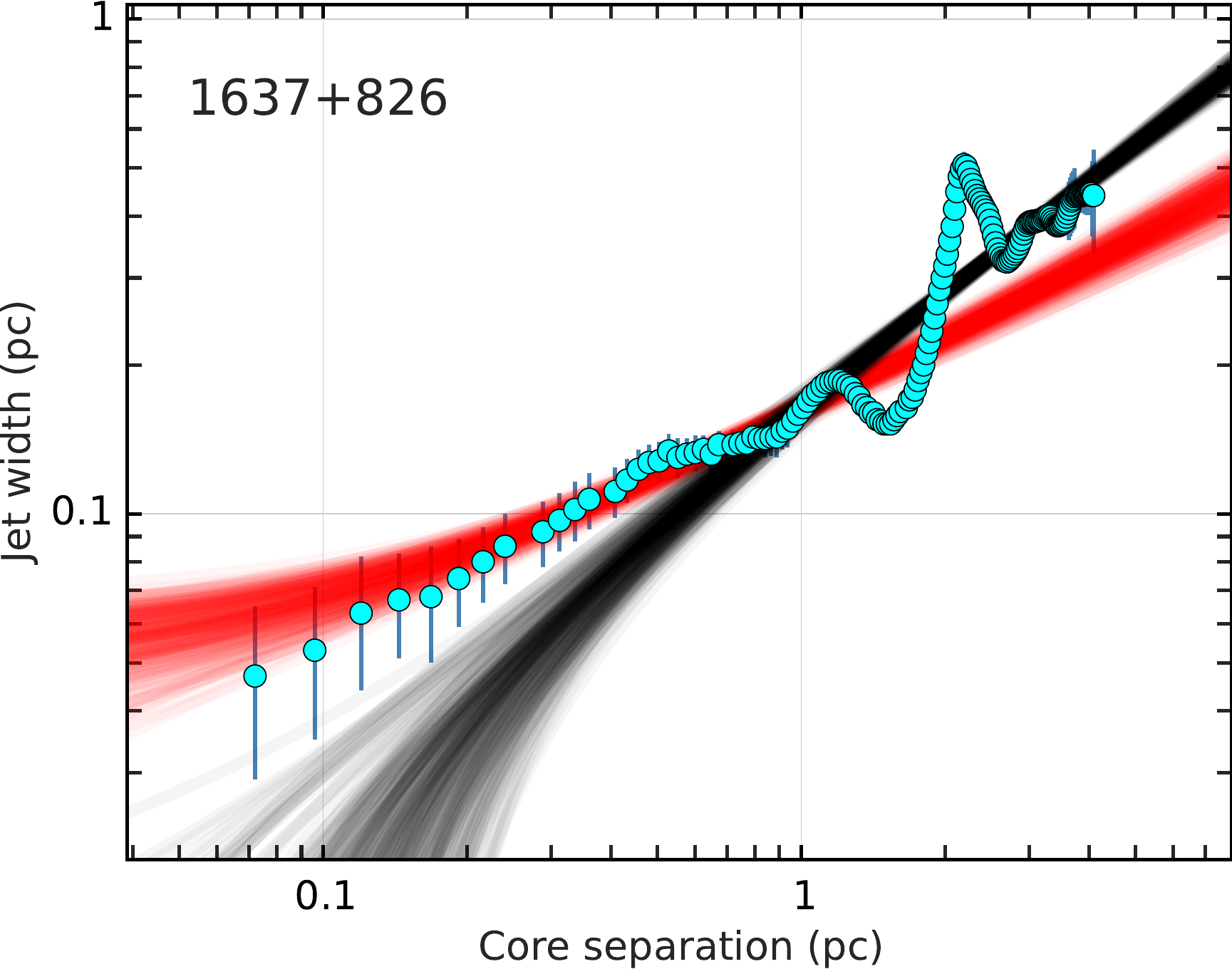}
\includegraphics[width=0.33\textwidth]{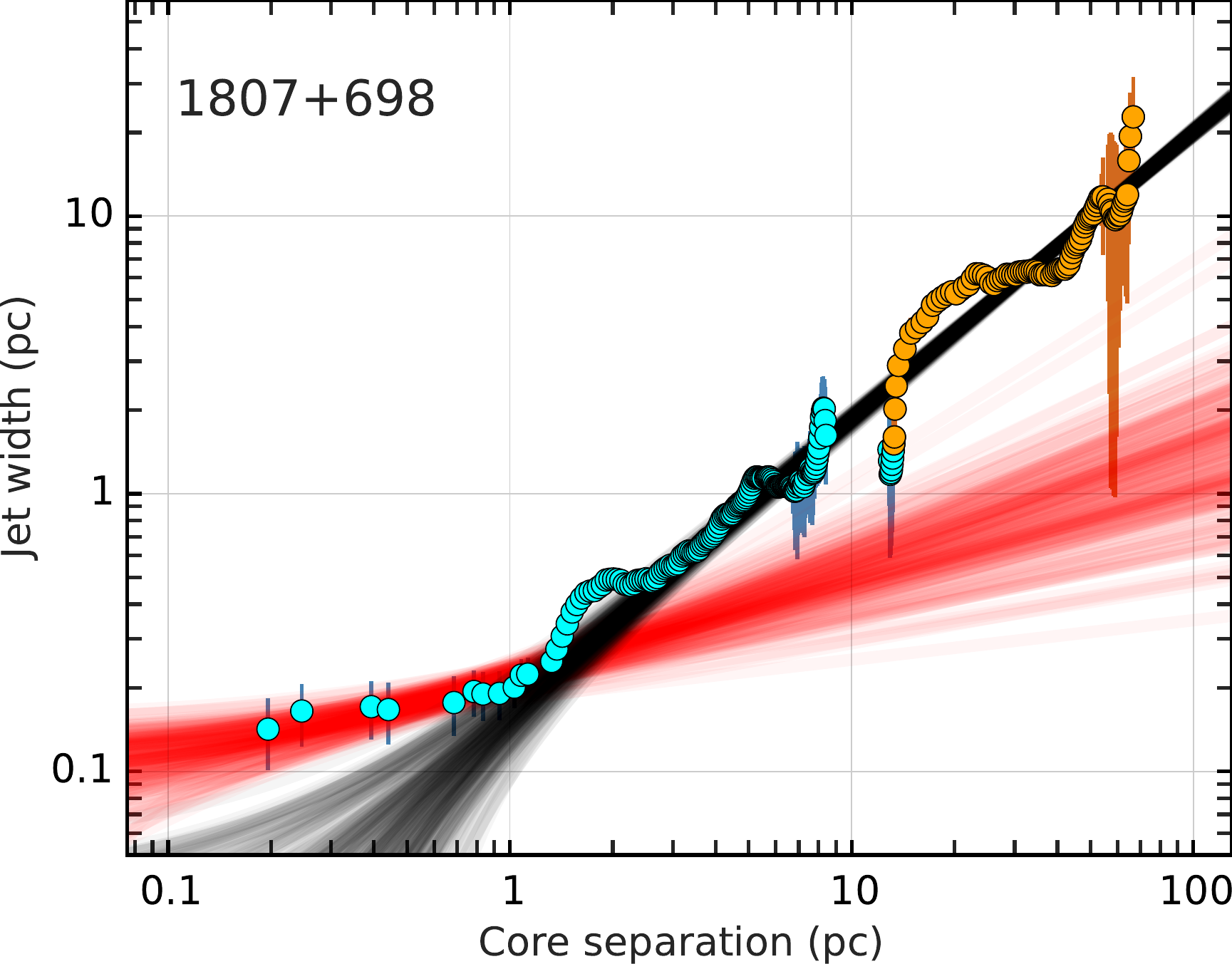}
\vskip 4mm
\includegraphics[width=0.33\textwidth]{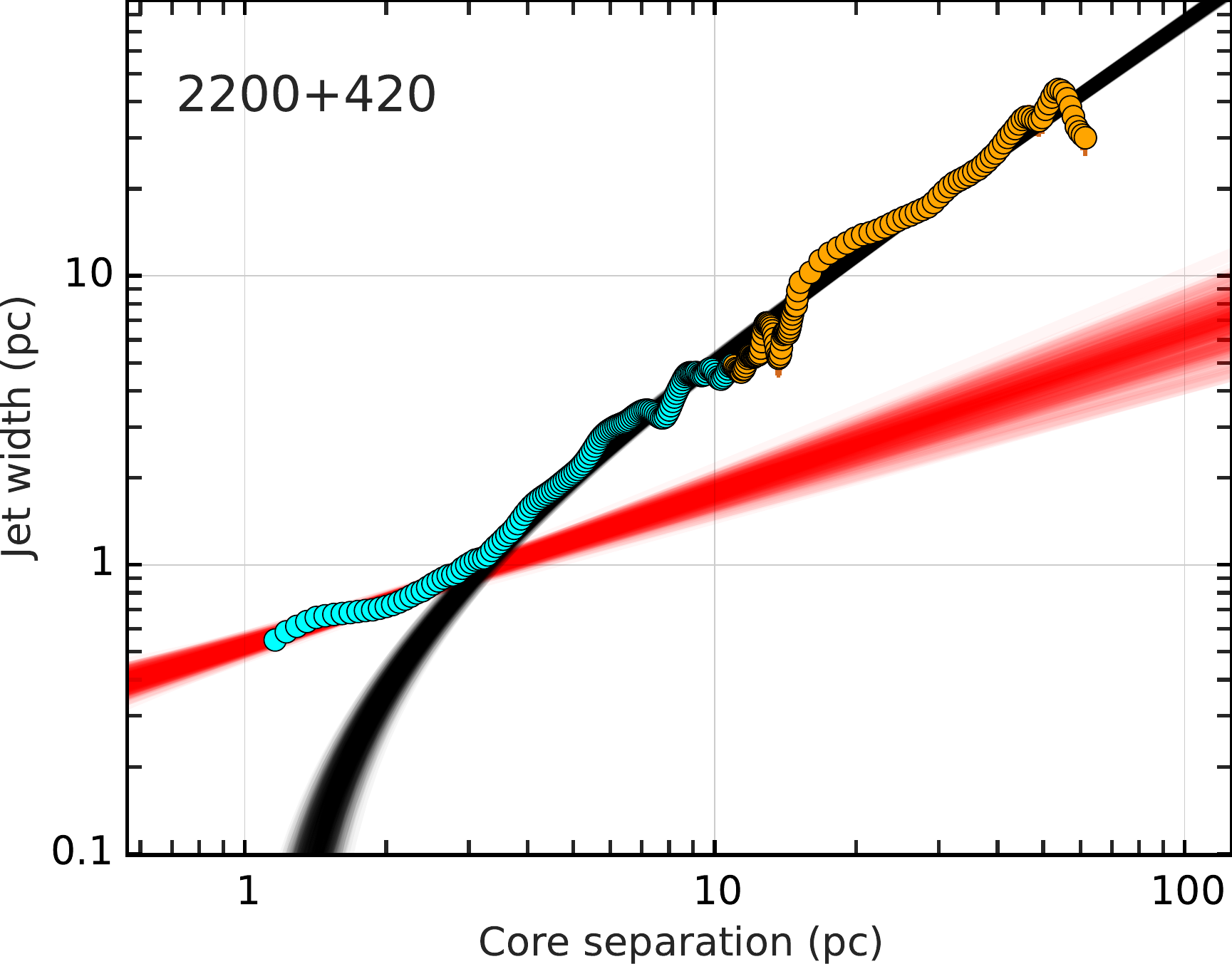}
\vskip 2mm
\caption{
Jet profiles with an indication of transition from parabolic to conical shape in ten well resolved nearby active galaxies.
The dependence of the jet width on projected distance from the apparent jet base is shown.
The cyan and orange dots show measurements at 15~GHz and 1.4~GHz, respectively.
The red and black stripes represent Monte Carlo fits for jet regions before and beyond the jet shape transition region, respectively.
The projected distance is shown in pc for targets with known redshift and in mas for 0815$-$094 which has no redshift information.
General properties of these AGN are presented in \autoref{t:sourprop},
parameters of the fits --- in \autoref{t:mc_twofits_pars}, 
parameters of the shape transition region --- in \autoref{t:break}.
\label{f:geometry_transition}
}
\end{figure*}

\setcounter{table}{1}
\begin{table*}
\caption{Derived best-fit parameters of the two fitted dependencies $d=a_1(r+r_0)^{k_1}$ and $d=a_2(r+r_1)^{k_2}$ before and after the jet break, respectively (\autoref{f:geometry_transition}). We used the VLBA data at 15~GHz only (band `U') or 15~GHz and 1.4~GHz (band `UL') between $r_\mathrm{min}$ and $r_\mathrm{max}$ distance from the apparent core.
Note that all the values of $r$ are projected on the plane of the sky.
}
\label{t:mc_twofits_pars}
\begin{tabular}{lccrccccc}
\hline\hline
Source         & Band & $r_{\rm min}$ & $r_{\rm max}$ & \multicolumn{2}{c}{$a_1$}        & \multicolumn{2}{c}{$r_0$} & $k_1$ \\
               &      &         (mas) &         (mas) & (pc$^{1-k_1}$) & (mas$^{1-k_1}$) &     (pc)    &      (mas)  &       \\
\hline
0111$+$021 &  U   & 0.2 &  1.5 & $0.179 \pm 0.010$   & $0.188 \pm 0.011$ & $0.143 \pm 0.085$  &  $0.157 \pm 0.093$  &  $0.497 \pm 0.077$ \\ 
0238$-$084 &  U   & 0.3 &  2.5 & $0.078 \pm 0.007$   & $0.319 \pm 0.046$ & $0.074 \pm 0.038$  &  $0.740 \pm 0.380$  &  $0.391 \pm 0.048$ \\ 
0415$+$379 &  U   & 0.2 &  6.0 & $0.305 \pm 0.011$   & $0.313 \pm 0.011$ & $0.042 \pm 0.020$  &  $0.044 \pm 0.021$  &  $0.468 \pm 0.026$ \\ 
0430$+$052 &  U   & 0.2 &  2.0 & $0.202 \pm 0.015$   & $0.245 \pm 0.020$ & $0.122 \pm 0.071$  &  $0.188 \pm 0.109$  &  $0.556 \pm 0.070$ \\ 
0815$-$094 &  U   & 0.2 &  1.0 &              \ldots & $0.294 \pm 0.015$ &            \ldots  &  $0.163 \pm 0.048$  &  $0.527 \pm 0.044$ \\ 
1133$+$704 &  U   & 0.3 &  1.5 & $0.437 \pm 0.013$   & $0.464 \pm 0.014$ & $0.061 \pm 0.046$  &  $0.069 \pm 0.052$  &  $0.528 \pm 0.040$ \\ 
1514$+$004 &  U   & 0.2 &  3.5 & $0.171 \pm 0.011$   & $0.171 \pm 0.011$ & $0.189 \pm 0.088$  &  $0.189 \pm 0.088$  &  $0.564 \pm 0.048$ \\ 
1637$+$826 &  U   & 0.2 &  3.0 & $0.155 \pm 0.005$   & $0.223 \pm 0.010$ & $0.098 \pm 0.044$  &  $0.204 \pm 0.092$  &  $0.506 \pm 0.041$ \\
1807$+$698 &  U   & 0.2 &  1.4 & $0.207 \pm 0.016$   & $0.210 \pm 0.016$ & $0.130 \pm 0.089$  &  $0.133 \pm 0.091$  &  $0.388 \pm 0.087$ \\ 
2200$+$420 &  U   & 0.9 &  2.0 & $0.505 \pm 0.029$   & $0.449 \pm 0.027$ & $0.087 \pm 0.096$  &  $0.067 \pm 0.074$  &  $0.537 \pm 0.057$ \\ 
\hline
Source         & Band & $r_{\rm min}$ & $r_{\rm max}$ & \multicolumn{2}{c}{$a_2$}        & \multicolumn{2}{c}{$r_1$} & $k_2$ \\
               &      &         (mas) &         (mas) & (pc$^{1-k_2}$) & (mas$^{1-k_2}$) &      (pc)   &      (mas)  &       \\
\hline
0111$+$021 &  U   & 1.5 &  14.0 & $0.252 \pm 0.031$   & $0.254 \pm 0.031$ & $-1.126 \pm 0.181$ & $-1.237 \pm 0.199$  &  $0.934 \pm 0.054$ \\
0238$-$084 &  U   & 2.5 &  14.1 & $0.252 \pm 0.021$   & $0.228 \pm 0.047$ & $-0.069 \pm 0.067$ & $-0.690 \pm 0.670$  &  $1.052 \pm 0.081$ \\
0415$+$379 &  UL  & 6.0 &  60.8 & $0.123 \pm 0.019$   & $0.122 \pm 0.019$ & $-2.043 \pm 0.100$ & $-2.151 \pm 0.105$  &  $1.175 \pm 0.046$ \\
0430$+$052 &  UL  & 2.0 & 121.9 & $0.229 \pm 0.022$   & $0.216 \pm 0.021$ & $-0.500 \pm 0.188$ & $-0.769 \pm 0.289$  &  $1.131 \pm 0.027$ \\
0815$-$094 &  U   & 1.0 &  14.3 &              \ldots & $0.282 \pm 0.033$ &             \ldots & $-0.085 \pm 0.097$  &  $1.032 \pm 0.049$ \\
1133$+$704 &  U   & 1.5 &   5.1 & $0.921 \pm 0.057$   & $0.941 \pm 0.058$ & $-0.753 \pm 0.083$ & $-0.857 \pm 0.094$  &  $0.828 \pm 0.047$ \\
1514$+$004 &  U   & 3.5 &  14.3 & $0.185 \pm 0.010$   & $0.185 \pm 0.010$ & $-1.167 \pm 0.102$ & $-1.167 \pm 0.102$  &  $0.886 \pm 0.022$ \\
1637$+$826 &  U   & 3.0 &   8.9 & $0.175 \pm 0.007$   & $0.213 \pm 0.010$ & $-0.089 \pm 0.048$ & $-0.185 \pm 0.100$  &  $0.730 \pm 0.029$ \\
1807$+$698 &  UL  & 1.4 &  85.2 & $0.179 \pm 0.018$   & $0.179 \pm 0.018$ & $-0.120 \pm 0.188$ & $-0.122 \pm 0.192$  &  $1.023 \pm 0.025$ \\
2200$+$420 &  UL  & 2.0 &  49.0 & $0.433 \pm 0.016$   & $0.447 \pm 0.016$ & $-1.142 \pm 0.093$ & $-0.885 \pm 0.072$  &  $1.124 \pm 0.009$ \\
\hline
\end{tabular}
\end{table*}

\begin{table*}
\caption{Derived best-fit parameters of a single fit dependence $d=a(r+r_0)^k$ for 319 AGN. Their $k$ values are presented in \autoref{f:k_hist}. We used the VLBA data at 15~GHz only (band `U') or 15~GHz and 1.4~GHz (band `UL') between $r_\mathrm{min}$ and $r_\mathrm{max}$ distance from the apparent core.
Full table is available online only, the first seven rows are shown here for guidance.}
\label{t:mc_onefit_pars}
\begin{tabular}{ccrrccc}
\hline\hline
Source     & Band & $r_{\rm min}$ & $r_{\rm max}$ &        $a$    & $r_0$ & $k$ \\
           &      &         (mas) &         (mas) & (mas$^{1-k}$) & (mas) &     \\
\hline
0003$+$380 &  U & 0.5 &  2.5 & $0.213 \pm 0.059$ & $\phantom{+}0.043 \pm 0.285$ & $1.064 \pm 0.218$ \\
0006$+$061 &  U & 0.5 &  2.8 & $0.245 \pm 0.070$ & $\phantom{+}0.205 \pm 0.246$ & $1.378 \pm 0.187$ \\
0007$+$106 &  U & 0.5 &  1.6 & $0.356 \pm 0.162$ & $\phantom{+}0.092 \pm 0.259$ & $2.110 \pm 0.489$ \\
0010$+$405 &  U & 0.5 &  7.0 & $0.220 \pm 0.051$ &           $-0.087 \pm 0.268$ & $0.804 \pm 0.190$ \\
0011$+$189 &  U & 0.5 & 11.7 & $0.266 \pm 0.059$ & $\phantom{+}0.204 \pm 0.323$ & $0.744 \pm 0.177$ \\
0015$-$054 &  U & 0.5 & 12.4 & $0.200 \pm 0.049$ & $\phantom{+}0.333 \pm 0.340$ & $0.814 \pm 0.193$ \\
0016$+$731 & UL & 0.5 & 38.2 & $0.675 \pm 0.065$ &           $-0.230 \pm 0.080$ & $0.867 \pm 0.030$ \\
\hline
\end{tabular}
\end{table*}

In a previous work \citep{MOJAVE_XIV}, we analyzed parsec-scale radio VLBI images of jets in 362 active galaxies from the MOJAVE program \citep{MOJAVE_XV}. This sample is dominated by compact radio bright blazars with a jet at a small angle to the line of sight and a typical redshift $z\approx1$. However, some low luminosity nearby radio galaxies were also included. \citet{MOJAVE_XIV} show that while the majority of resolved jets have a shape close to conical, a significant fraction of the sample has observed deviations. A systematic change in jet width profile has been noted by \citet{Hervet17}, who explain it by using a stratified jet model with a fast spine and slow but relatively powerful outer layer. In this paper, we investigate if this outcome is partly affected by the typical finite angular resolution of VLBI observations. We probe a possible dependence of the jet shape on the distance $r$ from the nucleus. Furthermore, we perform a systematic search for a possible transition from one jet shape to another on the basis of 15~GHz and 1.4~GHz VLBA images.

The observation of jets with a change from parabolic to conical shape may provide an instrument to probe the MHD acceleration mechanism models as well as the ambient medium conditions. The change in jet shape in M87 \citep{Asada12} is coincident with the stationary bright feature HST-1, which can be associated with the change in ambient pressure profile and appearance of a recollimation shock due to pressure drop and abrupt expansion. This interpretation is supported by the measurements of external medium pressure by \citet{RF15} almost down to the Bondi radius $r_\mathrm{B}=2GM/c_\mathrm{s}^2$ (sphere of influence), with an observed mass density profile $\rho\propto r^{-1}$ (here $c_\mathrm{s}$ is a sound speed). The recently observed jet shape in 1H~0323+342 \citep{Hada18} demonstrates a similar behavior. On the other hand, there are models predicting a jet shape transition for a single power law pressure profile. The analytical model by \citet{Lyu09} predicts the transition from parabolic to conical form for certain regimes, as well as quasi-oscillations in jet shape
in the conical domain. This solution has been applied to the reconstruction of the recollimation shock properties of M87 by \citet{GL17}, with a predicted total jet power on the order of $10^{43}$~erg/s. 
The recent semi-analytical results for the warm jet matching the ambient medium with a total electric current closed inside a jet by \citet{BCKN-17} predicts a change in a jet shape from parabolic to conical for the Bondi pressure profile $P\propto r^{-2}$. In this work we follow the latter model and consider the results for a warm outflow in more 
detail.

The structure of the paper is the following: \autoref{s:obs_break} presents our results of a search for the jet profile change from parabolic to conical in a large sample of AGN jets, we suggest a model and interpret our findings in \autoref{s:model}, a discussion is presented in \autoref{s:discussion}. We summarize our work in \autoref{s:summary}.
Throughout this paper we will use the term ``core''  as the apparent origin of AGN jets that commonly appears as the brightest feature in VLBI images of blazars \citep[e.g.,][]{Lobanov98_coreshift,Marscher08}. 
We adopt a cosmology with $\Omega_m=0.27$, $\Omega_\Lambda=0.73$ and $H_0=71$~km~s$^{-1}$~Mpc$^{-1}$ \citep{Komatsu09}.

\section{A discovery of shape transition as a common effect in AGN jets}
\label{s:obs_break}

\subsection{Automated search of candidates with a change in jet geometry}

For the purposes of our study, we made use of data at 15~GHz from the MOJAVE program, the 2~cm VLBA Survey, and the National Radio Astronomy Observatory (NRAO) data archive \citep{MOJAVE_XV} for those sources that have at least five VLBA observing epochs at 15~GHz between 1994 August 31 and 2016 December 26 inclusive.
We used the 15~GHz VLBA total intensity MOJAVE stacked epoch images supplemented by single epoch 1.4~GHz VLBA images to derive apparent jet widths, $d$, as a function of projected distance $r$ from the jet core, and determined jet shapes similar to \citet{MOJAVE_XIV}. In that work we fitted the $d$--$r$ dependence with a single power law $d\propto r^k$. The index is expected to be $k\approx0.5$ for a quasi-parabolic shape and 1.0 for a conical jet. 
We note that even single-epoch observations at 1.4~GHz adequately reproduce source morphology, i.e., effectively fill jet cross-section due to a steep spectrum of synchrotron emission of the outflow, with a typical spectral index $-0.7$ measured between 2 and 8~GHz \citep{puskov12} and $-1.0$ between 8 and 15~GHz \citep{MOJAVE_XI}, making the low-frequency observations sensitive enough to probe jet morphology at larger scales.
In our analysis we use the jet width measurements made at 1.4~GHz
only on large scales, not covered by the 15~GHz data. These scales are typically beyond 10~mas. This allows us to neglect the core shift effect \citep[e.g.,][]{Kovalev_cs_2008,Sokol_cs2011,core_shift_var}, which is expected to be about 1~mas between 1.4 and 15~GHz. Its value can not be easily derived since it requires simultaneous observations at different frequencies. As a result, the jet widths estimated at 15~GHz smoothly transition to those at 1.4~GHz.

We have carried out a similar analysis allowing for a change in the jet shape. 
Using all available data (15~GHz only or combined data set at 15~GHz and 1.4~GHz) for each source, we performed a double power law fit of the jet width as a function of distance, dividing the jet path length in a logarithmic scale by two parts in proportion of 1:1, 1:2 and 1:3 to search for cases when the fitted $k$-index at inner scales was $0.5\pm0.2$, while at outer scales it was $1.0\pm0.2$. After such cases were identified automatically, we tuned the fits by setting the distance of the transition region by eye. 

We ended up dropping 36 AGN jets from the original samle of 367 objects as having unsatisfactory fits caused by either 
(i) non-optimal ridge line reconstruction for jets with strong bending,
(ii) numerous large gaps in jet emission,
(iii) too short a jet length
(iv) low intensity regions not captured well by our jet width fitting.
This resulted in a sample comprising 331 AGN jets. 

As a result of this analysis, we found a shape transition in ten jets (\autoref{f:geometry_transition}, \autoref{t:sourprop}) out of 367 analyzed. 
We emphasize that all the AGNs with detected transition of the jet shape turned out to have low redshifts $z<0.07$, i.e., have a high linear resolution of 15~GHz VLBA observations --- better than 1~pc. This is highly unlikely to occur by chance and provides additional strong evidence that this result is not an observational artifact but a real effect.
See discussion of the rest of analyzed low redshift AGN in the sample in \autoref{ss:common}.
%
%
%
Among the ten sources, there is one, the radio galaxy 0238$-$084 (NGC~1052), that shows a two-sided jet morphology. For this object, we analyzed the approaching, brighter outflow propagating to north-east direction, determining the position of a virtual VLBI core using a kinematic-based minimization method described in \cite{VRK03}.


Following our discovery of the shape transition preferentially occurring in nearby AGN, we supplemented our initial AGN sample of 362 targets from \citet{MOJAVE_XIV} with stacked images of five more low-$z$ AGN which had five or more 15~GHz VLBA observing epochs after the Pushkarev et al.\ analysis was finished. These were: 0615$-$172, 1133+704, 1200+608, 1216+061,  1741+196. All the stacked images are available from the \mbox{MOJAVE} database\footnote{\url{http://www.physics.purdue.edu/MOJAVE/}}.

\subsection{Rigorous fitting of the jet shape}

For each of the 10 sources found to have a jet geometry transition, we fit the data with the following dependencies: $d=a_1(r+r_0)^{k_1}$ and $d=a_2(r+r_1)^{k_2}$, describing a jet shape before and after the break. Here $r_0$ is understood as the separation of the 15~GHz apparent core from the true jet origin due to the synchrotron opacity \citep[e.g.,][]{Lobanov98_coreshift,MOJAVE_IX}, while $r_1$ shows how much one underestimates the jet length if it is derived from the data only beyond the geometrical transition of the jet. We note that this approach is more accurate but more computationally intensive than that used by \citet{MOJAVE_XIV} and applied in the original selection of jet break candidates. It is needed in order to better fit for jet shape close to the apex.

We fit these dependencies with Bayesian modeling using the NUTS Markov Chain Monte Carlo sampler based on the gradient of the log posterior density. It was implemented in PYMC3 \citep{PyMC3}, which automatically accounts for uncertainties of all the parameters in further inferences. The best fit parameters are listed in \autoref{t:mc_twofits_pars}, showing that initially the jets are quasi-parabolic with $k_1$ close to 0.5, while beyond the break point region the outflow manifests a streamline close to conical, with $k_2\approx1$.
The location of the jet shape break given in \autoref{t:break} is estimated as the intersection point of these two $d-r$ dependencies.
Note that \autoref{t:break} includes results on the jet shape transition region for two more sources, 1H~0323+342 and M87 taken from \citet{Hada18} and \citet{nokhrina2019}, respectively.
We also note that the shown error of the deprojected position of the break is propagated from the fitting procedure, it does not include uncertainties on the viewing angle and black hole mass.

\begin{figure}
\centering
\includegraphics[width=\columnwidth]{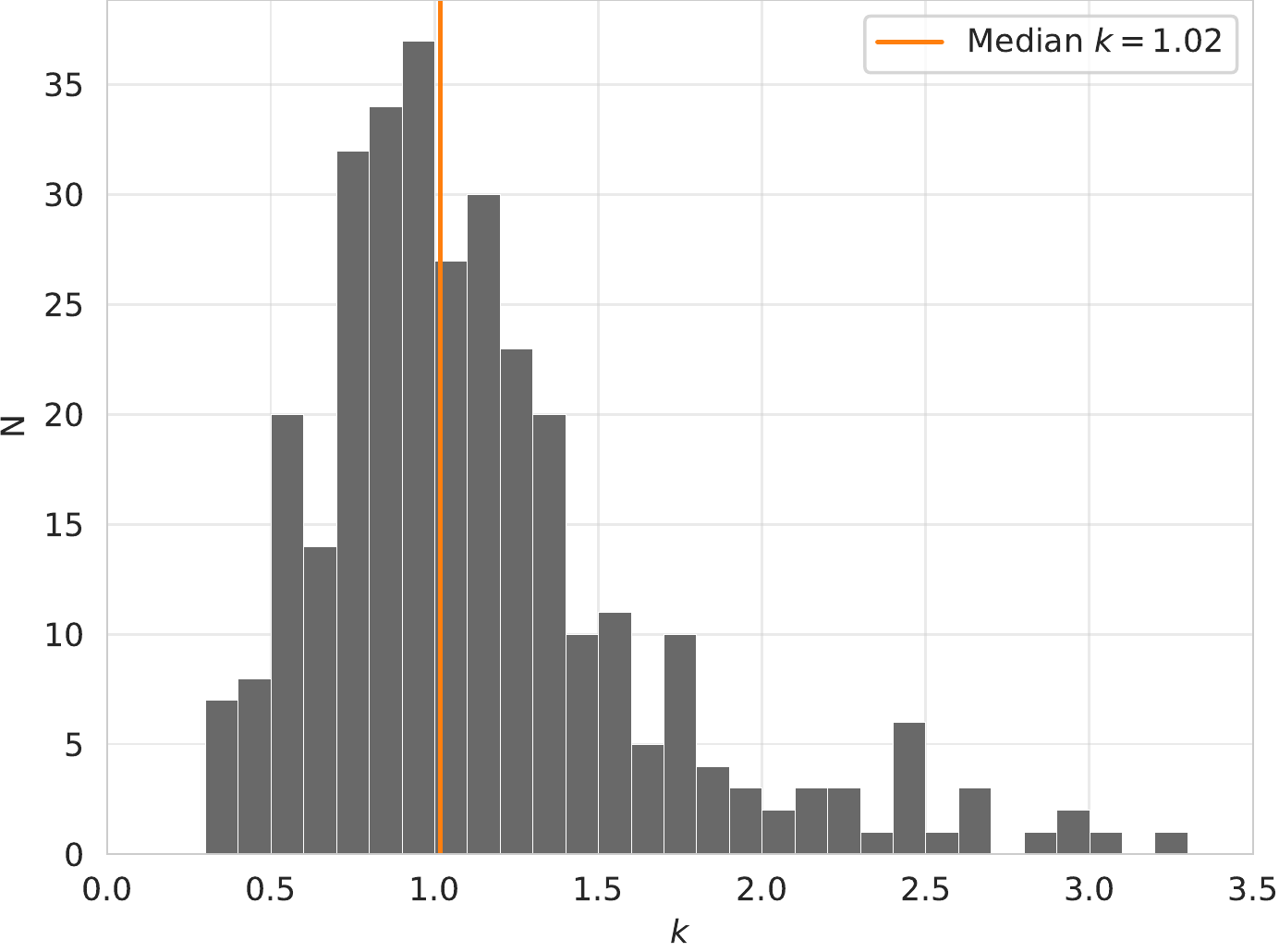}
\vskip -0.1cm
\caption{
A histogram of the best fit exponents $k$ assuming a single power law $d =a (r + r_0)^k$ for all spatial scales. Shown here are 319 sources from \autoref{t:mc_onefit_pars}.
\label{f:k_hist}
} 
\end{figure}

For other sources without a detected shape break, we fit a single power-law $d = a (r + r_0)^k$ for consistency. We excluded objects with unreliable ridge line detection or patchy structure in images (15 sources) and those with nonphysical $d-r$ dependence (24 sources) after visual inspection. They constitute only about one tenth of the dataset and thus the exclusion should not bias our estimates. To account for increased uncertainties of jet width measurements further from the core, the power law model is complemented as following:
$$
d = \begin{cases} a (r + r_0)^k + \mathcal{N}(0, \sigma_1^2), & \mbox{if } r < R \\ b + \mathcal{N}(0, \sigma_2^2), & \mbox{if } r > R \end{cases}
$$
Here all of $a, r_0, k, R, \sigma_1, b, \sigma_2$ are treated as unknown parameters and inferred simultaneously using a Nested Sampling algorithm as implemented in \texttt{PolyChord} \citep{polychord}. As expected, $\sigma_2$ is typically significantly larger than $\sigma_1$. We find that this model generally captures the $d-r$ dependence and its uncertainty well. Fitting results are given in \autoref{t:mc_onefit_pars} and the source distribution of exponents $k$ is shown in \autoref{f:k_hist}. Even though the estimates for individual sources have a large spread, the median exponent is very close to $1$. This indicates a conical average outflow shape, and agrees with previous results using slightly different estimation method \citep{pushkarev_etal12}.
We note the peak in the histogram bin at $k=0.5$ which corresponds to the parabolic jet shape; the number of objects with $k\approx0.5$ is not high enough in the sample to make it significant (\autoref{t:mc_onefit_pars}).

\subsection{Checking consistency of the fits and analyzing for possible biases}
\label{ss:consistency}

\begin{figure*}
\centering
\includegraphics[width=0.323\textwidth]{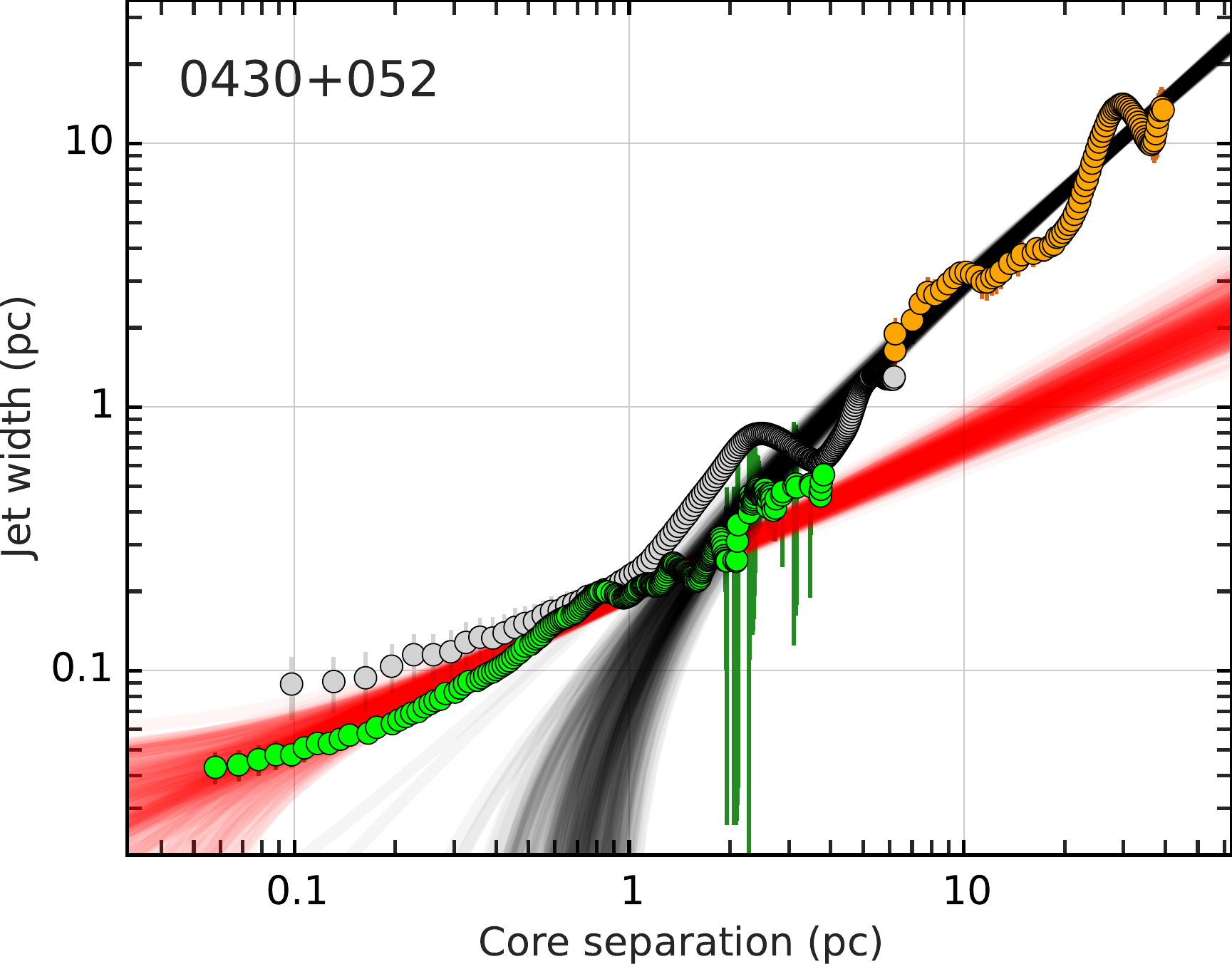}
~
\includegraphics[width=0.323\textwidth]{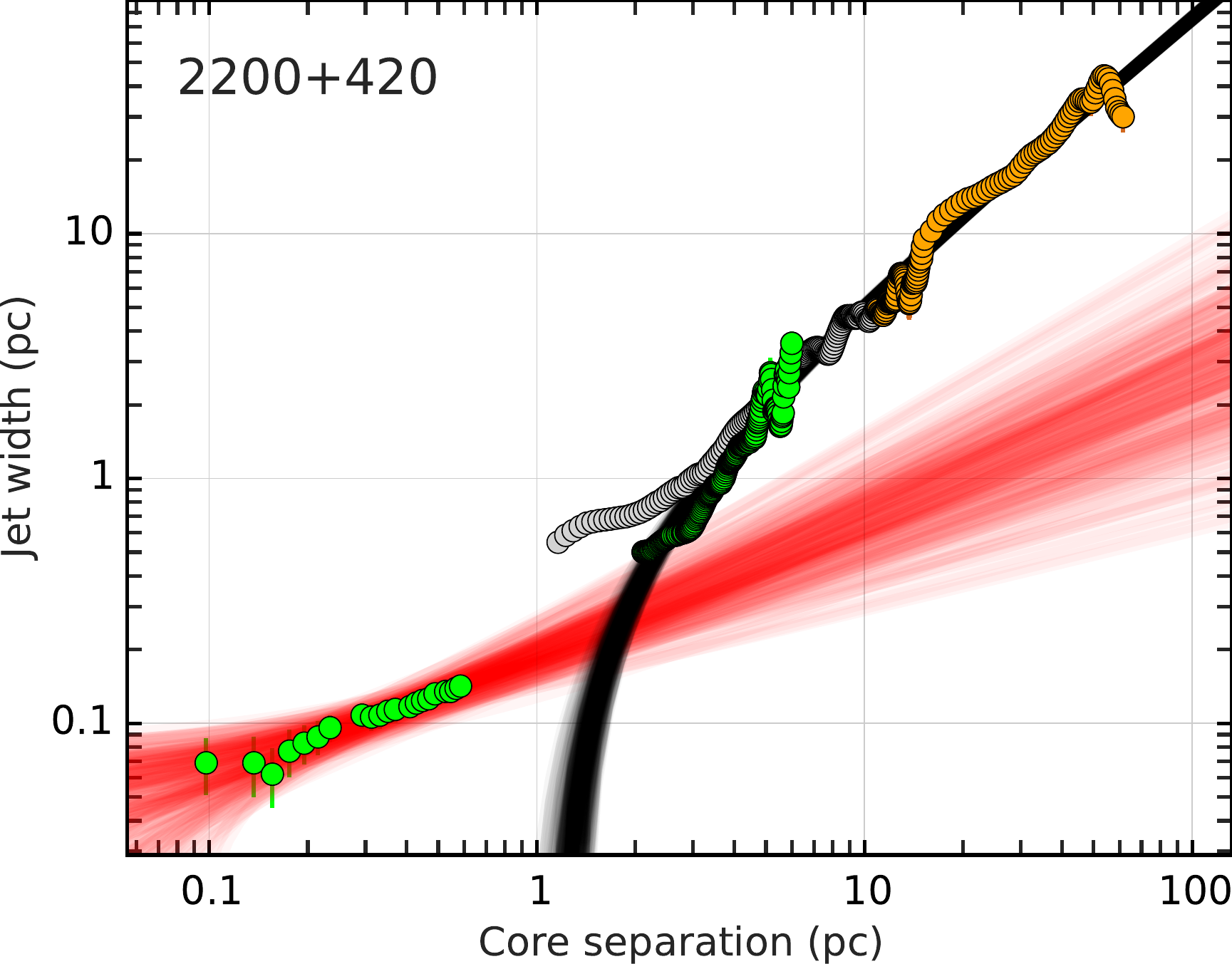}
\caption{
Dependence of the jet width on projected distance to apparent jet base for 0430+052 and 2200+420 in which 43~GHz data are used instead of 15~GHz.
The green and orange dots show measurements which are used in the fits at 43 and 1.4~GHz, respectively.
The 15~GHz measurements are not included in the fitting, they are shown by the grey color.
The red and black stripes represent Monte Carlo fits for jet regions before and beyond the jet shape transition region, respectively.
Parameters of the fits are as follows.
For 0430+052: $a_1 = 0.193\pm0.007$~pc$^{1-k_1}$,  $r_0 = 0.009\pm0.042$~pc, $k_1 = 0.586\pm0.047$, $a_2 = 0.241\pm0.026$~pc$^{1-k_2}$, $r_1 = -0.593\pm0.178$~pc, $k_2 = 1.116\pm0.029$.
For 2200+420: $a_1 = 0.187\pm0.025$~pc$^{1-k_1}$, $r_0 = 0.043\pm0.074$~pc, $k_1 = 0.571\pm0.097$, $a_2 = 0.430\pm0.018$~pc$^{1-k_2}$, $r_1 = -1.219\pm0.082$~pc, $k_2 = 1.126\pm0.011$.
Note that the derived $k$-values agree with 15~GHz results presented in \autoref{t:mc_twofits_pars}.
}
\label{f:geometry_transition_Q}
\end{figure*}

By setting $r=0$ we can estimate the apparent core size $d_\mathrm{c}^\mathrm{MC}$ at 15~GHz from the Monte-Carlo fit of the jet width as $a_1r_0^{k_1}$ and compare it with a median value of the core size $d_\mathrm{c}^{uv}$ derived from structure modelfit in visibility plane taken from \cite{MOJAVE_XVII}, see \autoref{t:core}. Two sources, 0111+021 and 0415+379, show a good agreement between $d_\mathrm{c}^\mathrm{MC}$ and $d_\mathrm{c}^{uv}$, while for the other seven objects $d_\mathrm{c}^\mathrm{MC}$ is somewhat larger than $d_\mathrm{c}^{uv}$. This is likely due to a non-ideal determination of the core position throughout the epochs, which is used to align single-epoch maps to produce stacked images. The radio galaxy 0430+052 is the only source having $d_\mathrm{c}^\mathrm{MC}<d_\mathrm{c}^{uv}$ for reasons that are unclear. 

A bias related to this effect might affect the results. Statistically analyzing jet shapes for the whole sample of 331 sources with stacked VLBA images by introducing different ridge line path length limits we have found the following. A near-parabolic streamline for quasars and BL~Lacs can be derived if the innermost jet, only up to $\sim$1~mas from the apparent core, is considered. This is not a real effect. 
The bias is found to be the most pronounced for curved jets or jets with features emerging at different position angles over time \citep{MOJAVE_X}. This is confirmed by an apparent artificial correlation of median jet width with the number of epochs in a stacked image for such AGN.
Uncertainties in the core position also contribute to this effect due to the imperfect alignment of images while performing the stacking. 
Variability of opacity conditions and apparent position of the core \citep{core_shift_var} affect this partially even though the alignment of the stacked single epoch images is done on the core position.
Together, it causes an additional artificial widening of the jet near the core region up to distances $r\approx0.3$~mas.
The effect quickly vanishes at larger scales. Thus, if we exclude jet width measurements at distances $\lesssim0.4$~mas, the effect becomes much weaker and disappears completely if we rule out the measurements within 0.5~mas from the core. We also note that radio galaxies, being at low redshift and thus having apparently wider outflows, are much less subject to this effect.
The same is true for the sources with a jet shape break shown in \autoref{f:geometry_transition}, as these are low-redshift objects. Only for BL~Lac, as the most remote source among them and also having a bright quasi-stationary component near the core \citep{Cohen14}, we put a conservative limit of 0.9~mas. For the other sources we used the non-cut intervals listed in \autoref{t:mc_twofits_pars}, because dropping measurements at $r<0.5$~mas did not significantly change the fit parameters. For the remaining sources, we have dropped all measurement for $r<0.5$~mas while analyzing the data (\autoref{t:mc_onefit_pars}).

Another possible problem might be related to cases where the jet width is completely unresolved. Indeed, this was found for some AGN targets at some epochs from the visibility model fitting of the core \citep[e.g.,][]{2cmPaperIV,MOJAVE_XVII}. We have addressed this issue by dropping all measurement for $r<0.5$~mas. Interestingly, the rest of the measured deconvolved jet width values are always positive. If we assume that this is some sort of a positive bias overestimating the width, it should not depend on $r$ for unresolved jets and will result in $k$ values close to zero. This behavior was not seen in our fitting results. 

We have also compared the fitted parameter $r_0$ with the core offset from the jet base estimated from the core shift measured between 15~GHz and 8~GHz \citep{MOJAVE_IX} assuming an inverse frequency dependence $r\propto\nu^{-1}$. These quantities, also listed in \autoref{t:core}, agree well within the errors in four out of six sources having measured core shifts. The large discrepancy for two sources can be explained by the recently recently established phenomenon of significant core shift variability \citep{core_shift_var} or the difference between the true jet shape derived by us and the assumed conical jet shape in \citep{MOJAVE_IX}.
We note that this result opens a new way to estimate the distance to the true jet origin which does not require an assumption regarding the jet geometry.

We checked and complemented our analysis using 43~GHz data from the Boston University (BU) AGN group\footnote{\url{https://www.bu.edu/blazars/VLBAproject.html}} for 0430+052 and 2200+420 (\autoref{f:geometry_transition}), which are present in both the MOJAVE and BU samples. For each of these sources we (i) produced stacked total intensity 43~GHz maps, aligning single epoch-images by the position of the VLBA core derived from structure modelfitting of the visibility data, (ii) determined the reconstructed jet ridge line, and (iii) fitted the transverse jet width as a function of distance from the core (\autoref{f:geometry_transition_Q}). It resulted in the same $k$-values before and after the break as in our original analysis within the errors (compare \autoref{f:geometry_transition_Q} and \autoref{t:mc_twofits_pars}). The jet shape transition region is found at core separations comparable to those from the 15~GHz data fits but has shifted slightly. 
We note however that 7~mm jet width estimates are systematically lower than those found from the 15~GHz data due to the weak high frequency synchrotron emission coming from the jet edges. Robust estimates of jet geometry and particularly of the jet width require high dynamic range images which are better sampled at intermediate radio frequencies. A good agreement between 15~GHz and 1.4~GHz width measurements increases the robustness of our results.

We warn readers about deriving jet shapes from structure model fitting of single-epoch data \citep[e.g.,][]{Hervet17}, as the jet may appear quasi-parabolic ($k<1$) up to a certain (typically short) distance from the core and then change its shape to conical ($k\approx1$). This effect occurs in the sources that show variations in their inner jet position angle. \citet{MOJAVE_X} established this as a common, decade-timescale phenomenon for the most heavily monitored AGNs in the MOJAVE sample. Thus, single-epoch VLBI maps may not reveal the whole jet cross-section, but rather a portion of it, especially in the inner jet regions where images are dynamic range limited. Therefore, the conclusions regarding jet geometry based strictly on a modelfit approach should be treated with caution.

\begin{table*}
\caption{Derived parameters of the jet shape break for 10 AGN with addition of 0321+340 adopted from \citet{Hada18} and 1228+126 from \citet{nokhrina2019}.
Columns are as follows: 
(1) source name (B1950); 
(2) jet width at the break in mas; 
(3) same as (2) but in pc; 
(4) projected distance of the break from the apparent core along the jet in mas; 
(5) projected distance of the break from the BH along the jet in mas; 
(6) same as (5) but in pc; 
(7) deprojected distance of the break from the BH along the jet in pc, the parameter uses the estimated viewing angle; 
(8) presence of a bright low pattern speed feature \citep{MOJAVE_XVII}.
\label{t:break}
}
\begin{tabular}{llllllrl}
\hline\hline
 Source         & \multicolumn{2}{c}{$d_{\rm break}$}                 & $r_{\rm break,\;app}^{\rm proj}$ & $r_{\rm break}^{\rm proj}$ & $r_{\rm break}^{\rm proj}$ & $r_{\rm break}^{\rm deproj}$ & Stationary \\
                & (mas)                               & (pc)          & (mas)                            & (mas)                      & (pc)                       & (pc)                         & jet feature \\
 (1)            & (2)                                 & (3)           & (4)                              & (5)                        & (6)                        & (7)                          & (8) \\
\hline
 0111$+$021     & $0.30\pm0.03$                       & $0.28\pm0.03$ & $2.46\pm0.27$                    & $2.62\pm0.29$              & $2.38\pm0.26$              & $27.31$                      & Y    \\
 0238$-$084     & $0.53\pm0.05$                       & $0.05\pm0.01$ & $2.93\pm0.57$                    & $3.73\pm0.65$              & $0.37\pm0.06$              & $0.49$                       & Y    \\
0321$+$340      & 1 & 1.16 & 10 & 10.04 & 11.64 & $106.07$ & Y$^a$ \\
 0415$+$379     & $0.78\pm0.03$                       & $0.74\pm0.03$ & $7.03\pm0.50$                    & $7.07\pm0.50$              & $6.72\pm0.47$              & $29.00$                      & \ldots\\
 0430$+$052     & $0.45\pm0.06$                       & $0.29\pm0.04$ & $2.67\pm0.40$                    & $2.85\pm0.41$              & $1.85\pm0.27$              & $5.77$                       & \ldots\\
 0815$-$094     & $0.37\pm0.05$                       & \ldots        & $1.37\pm0.30$                    & $1.54\pm0.30$              & \ldots                     & \ldots                       & \ldots\\
 1133$+$704     & $0.57\pm0.02$                       & $0.50\pm0.02$ & $1.39\pm0.09$                    & $1.46\pm0.10$              & $1.29\pm0.09$              & $14.80$                      & \ldots\\
1228$+$126      & $13.00\pm 0.50$ & $1.20\pm 0.04$ & \ldots & $131\pm 6$ & $10.50\pm 0.46$ & 43.41 & Y$^b$    \\
 1514$+$004     & $0.34\pm0.02$                       & $0.34\pm0.02$ & $3.10\pm0.22$                    & $3.39\pm0.30$              & $3.39\pm0.30$              & $13.10$                      & Y    \\
 1637$+$826     & $0.32\pm0.02$                       & $0.16\pm0.01$ & $1.92\pm0.26$                    & $2.13\pm0.28$              & $1.02\pm0.13$              & $3.30$                       & Y    \\
 1807$+$698     & $0.26\pm0.04$                       & $0.25\pm0.04$ & $1.53\pm0.33$                    & $1.67\pm0.34$              & $1.63\pm0.33$              & $12.83$                      & Y    \\
 2200$+$420     & $0.74\pm0.03$                       & $0.95\pm0.04$ & $2.45\pm0.10$                    & $2.52\pm0.13$              & $3.25\pm0.16$              & $24.57$                      & \ldots\\
\hline
\end{tabular}
\begin{flushleft}
$^a$From \cite{Hada18}.\\
$^b$From \cite{Asada12}. This refers to the well-known HST-1 feature \citep{M87_HST1_2cm} which is located too far downstream to be sampled by typical MOJAVE images.
\end{flushleft}
\end{table*}

\begin{table*}
\caption{Angular size of the VLBA core at 15~GHz, $d_\mathrm{c}^\mathrm{MC}$, and its offset from the true jet origin, $r_0^\text{MC}$, derived from our Monte Carlo modeling of the jet width compared with independent MOJAVE core size measurements in the visibility plane, $d_\mathrm{c}^{uv}$ \citep{MOJAVE_XVII}, and the core offset, $r_0^\text{cs}$, estimated from the multi-frequency core shift measurements \citep{MOJAVE_IX}.
The shown $d_\mathrm{c}^{uv}$ values are medians over all epochs from \citet{MOJAVE_XVII}.}
 \label{t:core}
 \begin{tabular}{ccccc}
 \hline\hline
    Source &  $d_\mathrm{c}^\mathrm{MC}$ &  $d_\mathrm{c}^{uv}$ &  $r_0^\text{MC}$ &  $r_0^\text{cs}$ \\
           &               (mas) &               (mas) &               (mas) &              (mas) \\
       (1) &                 (2) &                 (3) &                 (4) &                (5) \\
 \hline
0111$+$021 &  $0.075 \pm 0.025$  &  $0.079 \pm 0.030$  &  $0.157 \pm 0.093$  &  $0.159 \pm 0.050$ \\
0238$-$084 &  $0.282 \pm 0.067$  &  $0.284 \pm 0.042$  &  $0.740 \pm 0.380$  &             \ldots \\
0415$+$379 &  $0.073 \pm 0.017$  &  $0.075 \pm 0.008$  &  $0.044 \pm 0.021$  &  $0.275 \pm 0.050$ \\
0430$+$052 &  $0.096 \pm 0.034$  &  $0.182 \pm 0.014$  &  $0.188 \pm 0.109$  &  $0.051 \pm 0.050$ \\
0815$-$094 &  $0.113 \pm 0.020$  &  $0.062 \pm 0.041$  &  $0.163 \pm 0.048$  &             \ldots \\
1133$+$704 &  $0.116 \pm 0.049$  &  $0.089 \pm 0.048$  &  $0.072 \pm 0.056$  &             \ldots \\
1514$+$004 &  $0.067 \pm 0.018$  &  $0.043 \pm 0.027$  &  $0.189 \pm 0.088$  &             \ldots \\
1637$+$826 &  $0.100 \pm 0.025$  &  $0.069 \pm 0.004$  &  $0.204 \pm 0.092$  &  $0.198 \pm 0.050$ \\
1807$+$698 &  $0.096 \pm 0.031$  &  $0.067 \pm 0.007$  &  $0.133 \pm 0.091$  &  $0.240 \pm 0.050$ \\
2200$+$420 &  $0.105 \pm 0.064$  &  $0.044 \pm 0.003$  &  $0.067 \pm 0.074$  &  $0.090 \pm 0.050$ \\
\hline
\end{tabular}
\end{table*}

\begin{figure}
\centering
\includegraphics[width=\columnwidth, angle=0]{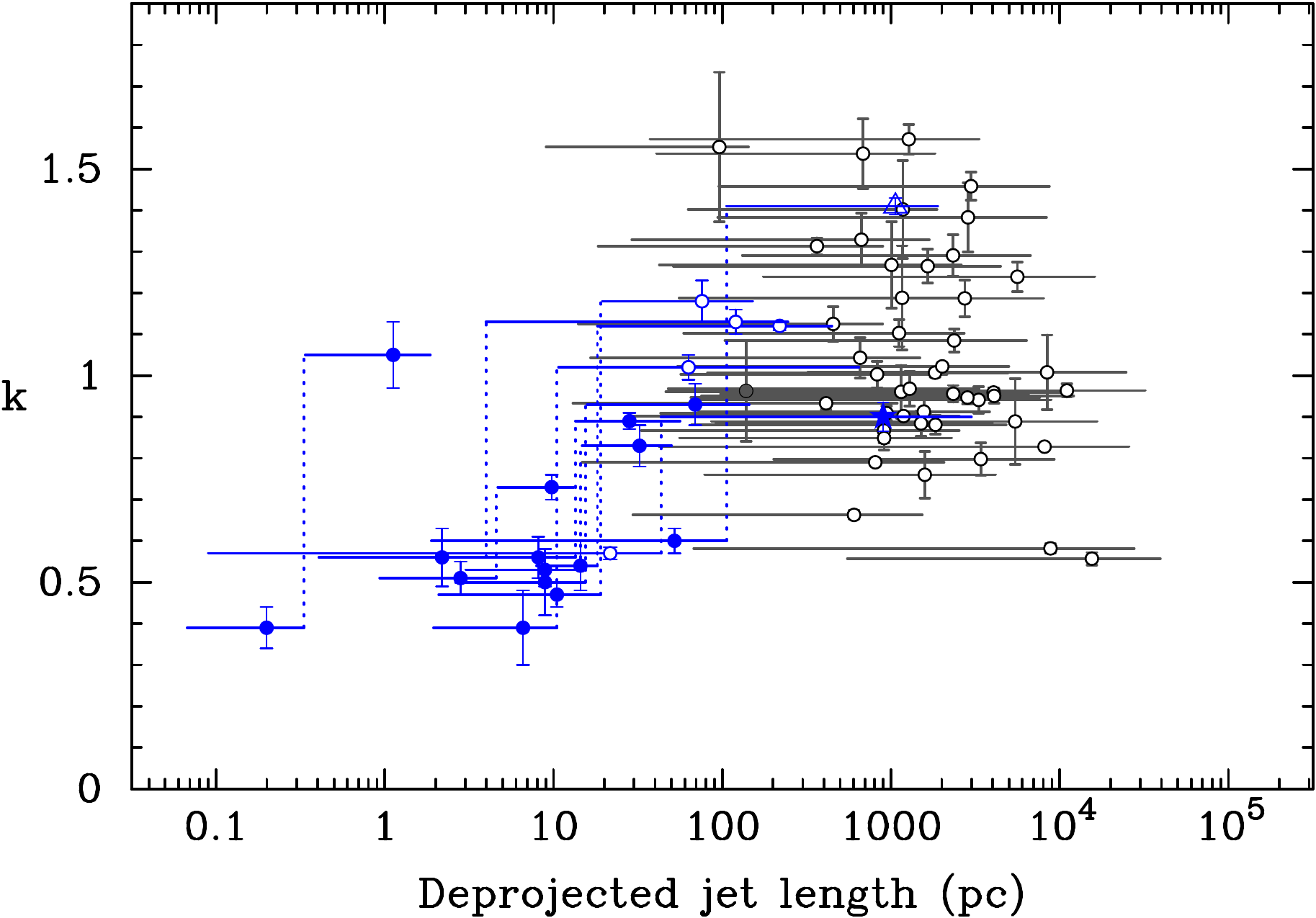}
\vskip -0.1cm
\caption{
Best fit $k$-index values plotted against deprojected distance from the 15~GHz VLBA core (\autoref{t:mc_twofits_pars}, \autoref{t:mc_onefit_pars}) for the sources listed in \autoref{t:sourprop} with measured redshift and viewing angle. 
Filled circles show fits at 15~GHz only, while empty circles denote results from analyzing measurements at 15~GHz and 1.4~GHz. 
Horizontal lines denote the scale over which the $k$-index was measured for every target. 
The symbols are placed at the median core distance of the analyzed jet portion. Eleven AGN with detected jet shape transition are shown in blue: 0111+021, 0238$-$084, 0321+340, 0415+379, 0430+052, 1133+704, 1514$+$004, 1637+826, 1807+698, 2200+420, and M87. The data for 0321+340 and M87 are taken from \citet{Hada18} and \citet{nokhrina2019}, respectively. 
\label{f:k_vs_r}
} 
\end{figure}

\subsection{Jet shape transition: a common effect in AGN jets, its consequences and prospects}
\label{ss:common}

We have found evidence for geometry transition in many jets for which sufficient linear resolution was achieved. This means that a change in jet shape is a common phenomenon which has significant consequences for many high angular resolution astrophysical and astrometric studies.
It is difficult to conclude if the geometry transition with measured properties is specific to only nearby radio galaxies and BL~Lacs, or can be extended to the AGN class in general. 
The radio luminosities of the nearby ($z<0.07$) jets are much lower than the rest of the sample and this might affect the geometry and transition zone.
We note that \autoref{f:k_vs_r} presents a consistent picture of the power index dependence on the downstream distance for nearby and distant jets.

In total, indications of the transition from parabolic to conical shape are found in 10 out of 29 nearby ($z<0.07$) jets observed as part of the MOJAVE program or by other investigators. VLBA archival data from the latter were processed by the MOJAVE team.
The reasons for non-detection of a geometry transition in nearby AGN jets are varied.  Some jets, e.g., 0007+106 and 1959+650, have too compact structure to study their shapes.  Some others, e.g., 0241+622, 0316+413, 1216+061, show purely parabolic streamlines (\autoref{t:mc_onefit_pars}), and their transition regions are expected at larger angular scales than those probed by our observations. E-MERLIN or low-frequency VLBA observations are needed. For example, the nearby radio galaxy 1216+061 ($z=0.0075$, scale factor 0.15~pc~mas$^{-1}$; not shown in \autoref{f:k_vs_r}) has a parabolic streamline with $k=0.64\pm0.05$ out to 7~mas at 15~GHz, corresponding to a deprojected distance of only $\approx1$~pc.
We are studying the remaining 12 low-redshift jets that show no sign of a profile break in a followup approved VLBA program.

The other jets in the sample (\autoref{t:mc_onefit_pars}), namely 97\,\%, do not show a clear significant change in jet geometry. 
We explain this by (i) a large scale factor of the order of 8~pc\,mas$^{-1}$ for a typical source in the sample at a redshift of $z\sim1$ and (ii) a small viewing angle typically about several degrees \citep{MOJAVE_XIV}. Jet power may also play a role, since the MOJAVE sample is flux-density limited and the AGN with $z> 0.1$ typically have jet luminosities $\sim 2$ orders of magnitude higher than the lower-redshift ones. 
The jets with a detected shape change have an average scaling factor of 0.7~pc\,mas$^{-1}$ and, on average, larger viewing angle since 6 out of 12 are radio galaxies.
Thus, if a transition region is located at a distance of a few tens of pc, it corresponds to a projected angular separation of $\la$1~mas from the apparent jet base at 15~GHz, which is comparable to the typical interferometric restoring beam size. VLBI observations at higher frequencies may be more effective in registering the jet shape transition, since they provide a better angular resolution and are less subject to opacity effects. This would probe scales closer to the jet apex and possible dependencies between acceleration zone extension and the maximum bulk Lorentz factor or jet power, as predicted by \citet{PC15}. 
On the other hand, the steep spectrum of the optically thin jet emission hinders the tracing of the jet for long distances.
The small viewing angles of the bright AGN jets set another limit on any jet shape investigation in the innermost parts. The streamline of an outflow can be studied down to distances at which the jet half-opening angle is still smaller than viewing angle. As shown by \citet{MOJAVE_XIV}, the intrinsic jet opening angle reaches values of a few degrees at scales of the order of 10 pc. This suggests that the jet shape transition phenomenon might be more effectively studied for nearby AGNs that are oriented at larger angles to the line of sight.
After considering all the points discussed above, we have begun a dedicated VLBA program in 2019 to search for geometry transitions in 61 AGN jets with $z<0.07$ from observations at 15~GHz and 1.4~GHz.

It is a challenging problem to estimate the consequences of this result on astrometry and astrophysics of AGN.
VLBI astrometry delivers the position of the true jet apex only if the opacity driven core shift is proportional to the frequency as $r\propto\nu^{-1}$ \citep{Porcas_cs2009}. However, this is expected only for conical jets and synchrotron opacity \citep{Lobanov98_coreshift}. A non-conical jet base results in an extension of the true jet length between the apex and the observed opaque core. This also produces somewhat larger VLBI-\textit{Gaia} offsets for AGN positions \citep{KPP17,r:Gaia5} than predicted by \citet{Kovalev_cs_2008}.

\subsection{Deprojected position of the jet break}
\label{ss:deprojpos}

We chose the MOJAVE-1 sample of 135 AGN \citep{MOJAVE_V} to perform a direct comparison with the 12 jets showing the breaks. Our reasoning is as follows. Most of MOJAVE-1 targets were observed by VLBA not only at multiple 15~GHz epochs but also in a single epoch at 1.4~GHz, which increases the jet distance probed by our analysis. In addition, VLBI measurements of the apparent kinematics $\beta_\mathrm{app}$ \citep{MOJAVE_XVII} and variability Doppler factor estimates $\delta$ \citep{Hovatta_etal09,LPA17} are available for a large fraction of the sample. We need this information to derive deprojected distance values. These requirements result in a sample of 65 sources (\autoref{t:sourprop}) described in \citet{MOJAVE_XIV}.

We derived viewing angle estimates through the relation 
$$
\theta = \arctan\frac{2\beta_\text{app}}{\beta^2_\text{app}+\delta_\text{var}^2-1}\,
$$
to convert the jet distance from angular projected to linear deprojected. 
Note that this assumes the same beaming parameters for the flux density variability and jet kinematics.
For $\beta_\mathrm{app}$ we used the fastest non-accelerating apparent jet speeds from the MOJAVE kinematic analysis. For 1H~0323+342 we use $\theta=6.3\degr$, based on the observed superluminal motion \citep{MOJAVE_XIII} assuming $\theta=(1+\beta_\mathrm{app})^{-0.5}=\gamma^{-1}$, which minimizes the required bulk Lorentz factor $\gamma$. The other possible viewing angle value for this target $\theta=4\degr$ is based on the variability time scale \citep{Hada18}.
For the BL~Lac objects 0111+021 and 1133+704 we assumed a viewing angle of $5\degr$, typical for this class of AGN \citep{Hovatta_etal09,Savolainen_etal10,MOJAVE_XIV,LPA17}. For the radio galaxy 1514+004 we assumed a viewing angle of $15\degr$ which is typical for this class of AGN in our sample. 

In \autoref{f:k_vs_r}, we plot the corresponding single power-law $k$-index values derived from the 15~GHz and 1.4~GHz VLBA data \citep{MOJAVE_XIV} versus deprojected distance from the 15~GHz VLBA core for 62 sources. There are eleven sources with known deprojected linear jet distance that have a jet shape transition (\autoref{f:geometry_transition}, \autoref{t:sourprop}). They are  shown by a pair of points each from the double power-law fits. The BL Lac object 0815$-$094 is not shown in Fig.~\ref{f:k_vs_r}, as it does not have a measured spectroscopic redshift.
Our results on jet shape transition (\autoref{t:mc_twofits_pars}, \autoref{t:break}, \autoref{f:k_vs_r}) are supplemented by multi-frequency data for M87 from \citet{Nakamura+18}, with $k_1=0.57$, $k_2=0.90$, and break point position obtained by \citet{nokhrina2019}. For M87 we adopt $\theta=14\degr$ \citep{Wang09}, consistent with more recent results by \citet{Mertens16}. For NLSy1 1H~0323+342 we use 1.4--2.3~GHz measurements from VLBA observations \citep{Hada18}, with $k_1=0.6$ and $k_2=1.41$, for which the jet shape break point position is estimated. 

Horizontal lines represent the scales at which $k$-indices were derived, starting from several tens of mas distance from the 15~GHz VLBA core (see \autoref{ss:consistency}) and up to distances limited by the sensitivity of our observations. The nearby jets, for which we are probing closer to the central engine, have low $k$ values and show a transition from quasi-parabolic values at small scales to quasi-conical at larger scales (\autoref{f:k_vs_r}). 
It is possible that at scales greater than $\sim100$~kpc, where jets become diffuse and disruptive, their geometry further changes from conical to hyperbolic, characterized by more rapid expansion \citep{Owen_etal00}.

\begin{figure}
\includegraphics[width=\columnwidth, angle=0]{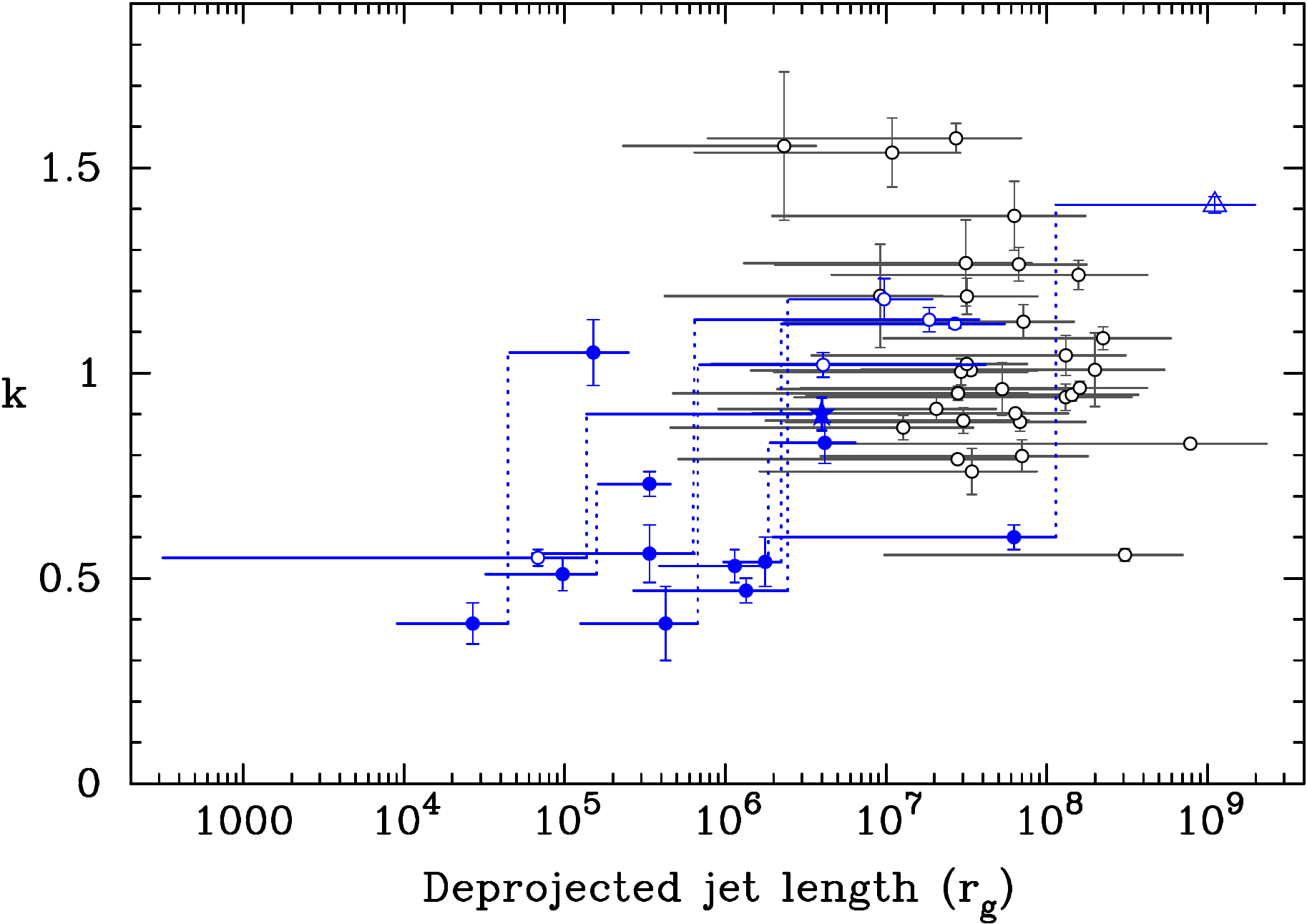}
\vskip -0.1cm
\caption{
Same as \autoref{f:k_vs_r} but with the deprojected distance measured in gravitational radius units. The black hole masses are listed in \autoref{t:sourprop}. When available, we use the mass estimates based on the velocity dispersion method, otherwise --- those from reverberation technique.
Note, the rightmost source with the detected transition from parabolic to conical shape is 1H~0323+342. Its mass estimate is based on reverberation mapping and might be strongly underestimated as argued by \citet{LeonTavares14} and \citet{Hada18}.
} 
\label{f:k_vs_rs}
\end{figure}

In order to plot the observed $k$-index values as a function of the deprojected distance along jets in gravitational radius $r_\mathrm{g}=GM/c^2$ units, we use the black hole masses estimated assuming virialized broad lines region (BLR) motion and correlation between BLR size and UV/optical luminosity \citep{Trrlb12, McLJ02, VP06, Landt17, Palma11, 2012ApJ...748...49S, Liu06}. We also use mass values inferred by stellar or gas ki\-ne\-ma\-tics methods \citep[e.g.,][]{WU02} for the closest sources. The mass values and references can be found in \autoref{t:sourprop}. We plot the data in \autoref{f:k_vs_rs}. It turns out that the sources with BH masses obtained by stellar velocity dispersion method or stellar/gas kinematics measurements are the subset of the sources with the detected jet shape break (i.e., the closest ones).

Since estimating the black hole mass is a complicated and strongly model-dependent method, some of the values might be significantly in error. By dropping the highest and lowest values as possible outliers of the derived jet break position $r_\mathrm{break}$ measured in $r_\mathrm{g}$ we are able to bound its values in the narrower range $r_\mathrm{break}\in(10^5,\,10^6)r_\mathrm{g}$.
This is an important result, especially when taken together with our finding that the jet shape transition may be a common phenomenon in nearby or even most of the AGN.

We note the following. The black hole mass of 1H~0323+342 is suspected to be underestimated \citep{LeonTavares14, Hada18}. If we use for this source the mass $M=10^{8.6}M_{\odot}$, obtained using the relation between black hole mass and bulge luminosity \citep{LeonTavares14}, 1H~0323+342 yields $r_\mathrm{break}=5.6\times 10^6r_\mathrm{g}$, falling much closer to the discussed above range of $r_\mathrm{break}/r_\mathrm{g}$ distances. This may provide an additional argument favoring a higher black hole mass for this source.

We have compared our results for the radio galaxy NGC\,6251 with those obtained earlier for this source by \cite{tseng16}. We have found that the jet shape transition region in this source is at $(1.6 \pm 0.2) \times 10^5\,r_\mathrm{g}$, assuming viewing angle of $18\degr$ and black hole mass of $6 \times 10^8\,M_\odot$ (see \autoref{t:sourprop}, \autoref{t:break}). This is slightly smaller compared to $(1-2) \times 10^5$ Schwarzschild radius estimated by \cite{tseng16}, who assumed the same black hole mass and a viewing angle of $19\degr$. The small difference might be caused by different techniques used to derive it. First, we measured transverse jet widths from the stacked image of the source, using 14 epochs at 15~GHz from the MOJAVE program and archival VLBA data. Second, we have taken into account the synchrotron opacity of the jet base by introducing the parameter $r_0$ that reflects an offset of the apparent 15~GHz core from the true jet apex. 

Of 12 sources with observed change in a jet boundary shape 6 are FR~I type, 2 are  FR~II type, and 4 have uncertain classification based on published radio images.
This may mean that different environments expected in these two different types of sources on large scales are either the same on the smaller scales, or affect the jet shape in the same way up to $10^6 r_\mathrm{g}$.


\section{Modeling relativistic jet with a shape break}
\label{s:model}

\subsection{Qualitative consideration}

Both analytical (see below) and phenomenological~\citep{PC13,PC15} considerations as well as 
numerical simulations~\citep{Komissarov_etal09, TMN09, Porth_etal11} show that for moderate initial magnetization 
of a jet $\sigma_{\rm M} \sim 10$--$10^2$, where 
\begin{equation}
\sigma_{\rm M} = \frac{\Omega_{0}^2 \Psi_{0}}{8 \pi^2 \mu\eta c^2}
\label{sigmaM}
\end{equation}
is the Michel magnetization parameter, the flow transits from a magnetically dominated regime at small distances $r$ from the origin to a particle dominated regime at larger distances. Here $\Psi_{0}$  and $\Omega_{0}$ are the total magnetic flux and characteristic angular velocity of the ``central engine''
respectively. Accordingly, $\mu = m_{\rm p}c^2 + m_{\rm p} w$ is the relativistic enthalpy, where $w$ is 
the non relativistic enthalpy, and $m_{\rm p}$ is a particle mass. Here we assume a leptonic jet, so $m_{\rm p}$ is the electron mass. Below for simplicity we consider not so large temperatures, so that
$w \ll c^2$. Finally, $\eta$ is the particle-to-magnetic flux ratio.

Indeed, the physical meaning of the Michel magnetization parameter is the maximum Lorentz factor $\gamma$ 
of the hydrodynamical flow when all the electromagnetic energy flux is transferred to particles. On the 
other hand, for quasi-cylindrical jets the following asymptotic solution for magnetically 
dominated flow exists \citep[see e.g.,][]{BeskinBook}
\begin{equation}
\gamma(r_{\perp}) = \frac{r_{\perp}}{R_{\rm L}},
\label{Gamma}
\end{equation}
where $R_{\rm L} = c/\Omega_{0}$ is the light cylinder radius, and  $r_{\perp}$ is the distance 
from the jet axis. For the black hole spin  $a_*=0.5$, $R_{\rm L} \approx 14.9\,r_{\rm g}\approx 2.2\times 10^{15}\;(M_{\rm BH}/10^9\,M_{\odot})\;{\rm cm}\approx 7.1\times 10^{-4}\;(M_{\rm BH}/10^9\,M_{\odot})\;{\rm pc}$. {Here and below we use the maximum BH energy extraction rate condition
$\Omega_{\rm F}=\Omega_{\rm H}/2$ \citep{BZ-77}.}
For observed pc scale jets, the jet width $d$ at the jet shape break point reaches 1 pc. This means that at the transition point $d/2R_{\rm L} > \sigma_{\rm M}$, and the flow cannot be still magnetically dominated. 
As was shown by~\citet{NBKZ15} who have analysed about 100 AGN jets, $\sigma_{\rm M} \sim 10-50$ is a reasonable 
value constrained by the observations. 
The observed median value of 1.02 for the $k$-index also clearly points to a ballistic plasma motion. This suggests that the jet is dominated by the plasma bulk motion kinetic energy at the deprojected distance longer than $\sim100$~pc or $\sim10^7 r_\mathrm{g}$ rather than by the Poynting flux, as expected close to the launching region.  

For this reason we aim to explain the break in the $d(r)$ dependence as a consequence of 
a transition from the magnetically dominated to the particle dominated regime. Below we present the main 
results of our semi-analytical consideration. Our goal is in evaluating the dependence of the jet width $d$ on an ambient pressure profile $P_{\rm ext}(r)$. The results for the cold jet are presented in 
\citet{BCKN-17}, while here we consider the semi-analytical results for a warm outflow.

\subsection{Semi-analytical model}
\label{ss:Model}

Basic equations describing the internal structure of relativistic and non relativistic jets within the Grad-Shafranov approach
are now well-established~\citep{HN89, PP92, LHAN98, BM00, BeskinBook, Lyu09}. 
This approach allows us to formulate the problem of finding a
stationary axisymmetric magnetohydrodynamic outflow structure (a jet solution) using a set of two differential equations on a magnetic flux function $\Psi$ and an Alfv\'enic Mach number ${\cal M}$. These equations are Bernoulli equation and Grad--Shafranov equation of a force balance perpendicular to magnetic surfaces.  
The approach allows us to
determine the internal structure of axisymmetric stationary jets knowing in general case five ``integrals of motion'', 
i.e., energy $E(\Psi)$ and angular momentum $L(\Psi)$ flux, electric potential which connects with angular velocity
$\Omega_{\rm F}(\Psi)$, entropy $s(\Psi)$, and the particle-to-magnetic flux ratio $\eta(\Psi)$. 
All these values are to be constant along magnetic surfaces $\Psi = \mathrm{const}$. Once the Grad--Shafranov and Bernoulli equations are solved for the given integrals, all the other flow properties, such as particle number density, four-velocity, electric current, and Lorentz factor, can be determined from algebraic equations \citep[e.g.,][]{BeskinBook}. In particular, it was 
shown that a jet with total zero electric current can exist only in the presence of an external medium with 
non-negligible pressure $P_{\rm ext}$. Thus, it is the ambient pressure $P_{\rm ext}$ that is expected to determine the 
transverse dimension of astrophysical jets.
In general, it is a complicated problem to solve the set of Bernoulli and Grad--Shafranov equations. An additional complication is connected with the change of a system type from elliptical to hyperbolic. So, to tackle the problem different simplifications are introduced. Here we simplify the problem, assuming the flow is highly collimated and can be described within the cylindrical geometry, in which case it can be solved numerically \citep{BM00}. 

On the other hand, careful matching of a solution inside the jet with the external medium has not been achieved 
up to now. The difficulty arises with having a very low energy density of the external medium in comparison with 
the energy density inside the relativistic jet. For this reason, in most cases an infinitely thin current 
sheet was introduced. Moreover, an ambient pressure was often modelled by homogeneous magnetic field 
$B_{\rm ext}^2/8 \pi = P_{\rm ext}$.

Below we use the approach developed by \citet{BCKN-17}. This paper is  later referred to as B17. We propose a flow with an electric current closing fully inside a jet. This is achieved by a natural assumption that the integrals $L$ and $\Omega_{\rm F}$ vanish at the jet boundary. The second assumption of the model is a vanishing flow velocity at the jet boundary, which leads to vanishing of a poloidal magnetic field component along with a toroidal due to current closure. As a consequence, only a thermal pressure, defined by a sound velocity $c_{\rm jet}$ and particle number density $n_{\rm jet}$, is left at the jet boundary to balance the ambient medium pressure without a current sheet. We solve Grad--Shafranov and Bernoulli equations for the flux function $\Psi(r_{\perp})$ and the square of an Alfv\'enic Mach number ${\cal M}^2(r_{\perp})$. 
The local non-relativistic enthalpy $w$ for a polytropic equation of state with politropic index $\Gamma=5/3$ can be written as
\begin{equation}
w = \frac{c_{\rm jet}^2}{(\Gamma - 1)} 
\left(\frac{n}{n_{\rm jet}}\right)^{\Gamma - 1},
\label{wM}
\end{equation}
where the local particle number density $n$ is obtained from the equation
\begin{equation}
{\cal{M}}^2n=4\pi\eta^2m_{\rm p}c^2\left[1+\frac{1}{\Gamma-1}\frac{c_{\rm jet}^2}{c^2}\left(\frac{n}{n_{\rm jet}}\right)^{\Gamma-1}\right].\label{neq}
\end{equation}

We solve the system of MHD equations (B17) for the boundary conditions $\Psi(0)=0$ and 
\begin{equation}
\left.P\right|_{r_{\perp}=d/2-0}=P_{\rm ext}.\label{second_bc}
\end{equation}
We should note that due to vanishing of the integrals $L(\Psi)$ 
and $\Omega_{\rm F}(\Psi)$ at the jet boundary, the thickness of the final current closure domain tends to zero and in B17 it was is not resolved. 
However, as it was shown, that the total pressure in this region is strictly conserved:
\begin{equation}
\frac{d}{dr_{\perp}}\left(P+\frac{B^2}{8\pi}\right)=0.\label{Pconst}    
\end{equation} 
This means that the solution we obtain up to the boundary does contain
the residual current and, thus, the toroidal magnetic field $B_{\varphi}$.

The main difference between the result presented here and the result by B17 is in more accurate account for the thermal terms, which can be seen in \autoref{neq}. To obtain the solution, we employ the following iterative procedure.
For each fixed fast magnetosonic 
Mach number at the axis ${\cal{M}}_0^2$ we initially set $P_{\rm ext}$ at the jet boundary. It defines the particle number density at the boundary $n_{\rm jet}$, and together with ${\cal{M}}_0^2$ --- the particle number density at the axis $n_0$. Having set the latter, we solve MHD equations across a jet from the axis outwards and calculate the jet pressure at the boundary provided by the solution $P^\mathrm{(solution)}$. By iterations we find self-consistently such $P_\mathrm{ext}$ that is equal to one, provided by the solution: $P^\mathrm{(solution)}=P_\mathrm{ext}$. Thus, we obtain the dependence of a jet pressure at the boundary as a function of a local jet width $d$.

This procedure fully determines the solution of our problem. 
For each magnitude of the external pressure the obtained solution is a crosscut at $r={\rm const}$.
Piling of these different crosscuts is a solution for an outflow in which one may neglect by
the derivatives over $r$ in comparison with the derivatives over $r_{\perp}$ in the two-dimensional
Grad--Shafranov and Bernoulli equations. This can be done for highly collimated, at least
as a parabola, outflows \citep{NBKZ15} and flows with small opening angles \citep{TMN09}.

We find that for the chosen sound velocity at the boundary $c_0^2=0.001 c^2$ the thermal
effects may be neglected in the outflow volume, playing an important role only
at the outflow boundary. It turns out that the resultant dependence of pressure at the jet boundary as a function of jet radius obtained by B17 and here start to differ somewhat only for large ${\cal{M}}_0^2$ (this value is of an order of 10, but depends on the initial magnetization), affecting the flow boundary shape downstream of the equipartition transition, and the effect on $k_2$-index is of the order of a few per cent. We will address the particular effects of higher temperature in the future work.

The proposed jet model with an electric current enclosed inside the jet
has a natural sheath structure, observed, for example, in the M87 jet \citep{Mertens16}. Due to choice of integrals, the outer parts of a jet have slower velocities, tending to non-relativistic with $\gamma(d/2)=1$. Such a sheath may be produced by different mechanisms: it may be 
a slower disk wind or an outer jet disturbed and slowed down by the pinch instability \citep{Tchekhovskoy19}. In our model it appears naturally as a consequence of a jet transiting into the ambient medium with the hydrodynamical discontinuity only (B17). 

\subsection{Transition from magnetically dominated to particle dominated flow}
\label{ss:transition}

\begin{figure}
\centering
\includegraphics[width=\columnwidth, angle=0]{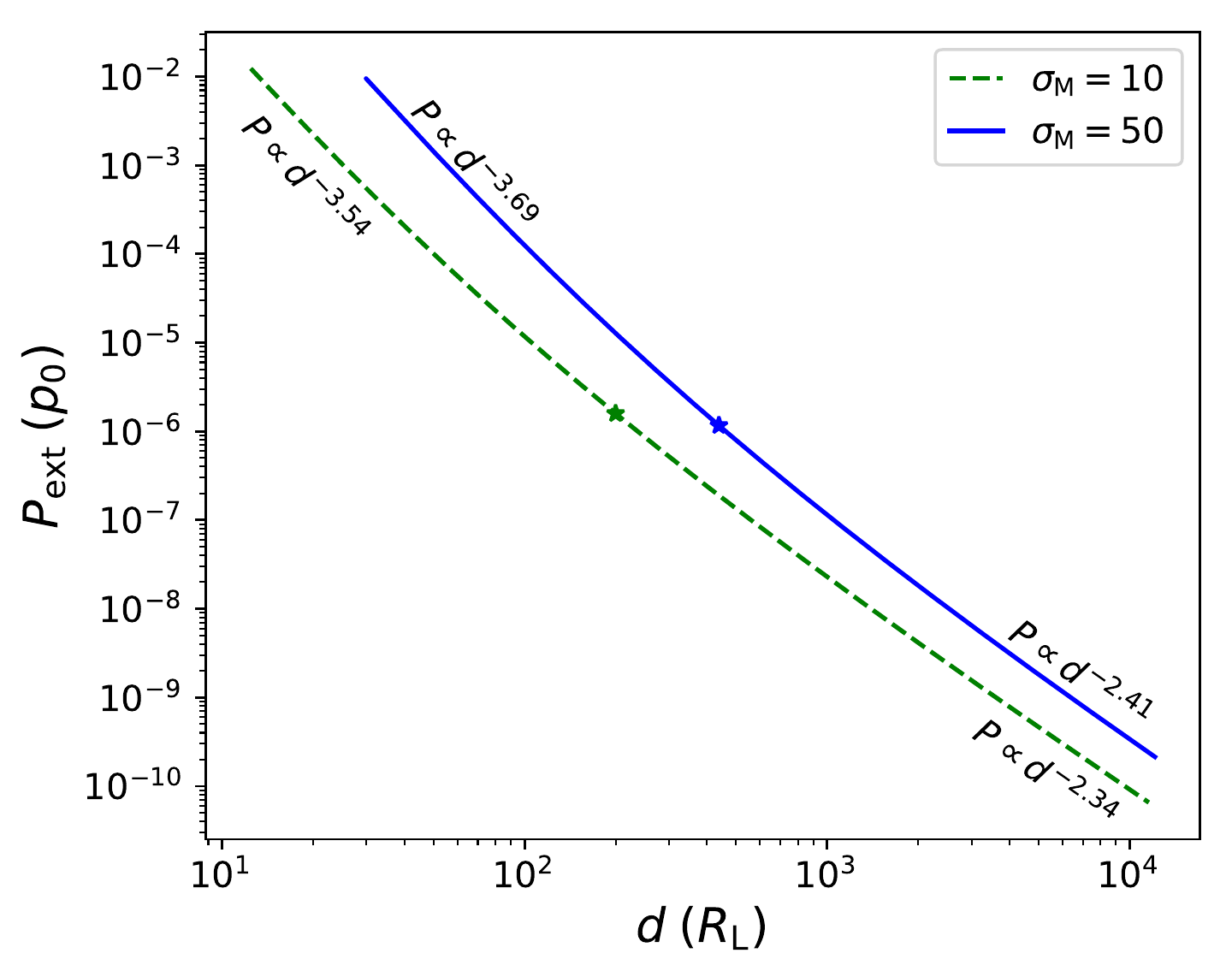}
\vskip -0.2cm
\caption{
Dimensionless ambient pressure $\tilde{p}$ as a function of a dimensionless jet width $\tilde{d}$ 
for the different magnetization parameters $\sigma_{\rm M}=10,\,50$, with log slopes inicated in the figure.
The start of each curve 
corresponds to the start of a super Alfv\'enic flow down from the axis.
The position of a break in the power law slope, designated by a star, depends on the flow initial magnetization in correspondence with our interpretation.  
}
\label{plot3}
\end{figure}

It is necessary to stress that this system of equations can describe both magnetically 
and particle dominated flow, with the physical answer (including the jet boundary radius $d/2$) 
depending on one external parameter only, namely, on the ambient pressure $P_{\rm ext}$.
In \autoref{plot3} we show the dependence of the dimensionless ambient pressure $\tilde{p}$ on a dimensionless 
jet width $\tilde{d}$ obtained by solving numerically the system of Grad--Shafranov and Bernoulli equations B17. The pressure is plotted in units of 
\begin{equation}
p_0=\left(\frac{\Psi_0}{2\pi R_{\rm L}^2\sigma_{\rm M}}\right)^2,\label{p0}
\end{equation}
so that $P_{\rm ext}=\tilde{p}\,p_0$, and the jet width
in
units of light cylinder radius is $d=\tilde{d}\,R_{\rm L}$.
We observe (see \autoref{plot3}) that the pressure
has a different power law  dependence on the jet radius for small and large $d$. For each magnetization $\sigma_{\rm M}$, this behavior holds, with the change between two profiles occurring at different jet widths. For $\sigma_{\rm M}=50$  the pressure changes its dependence on $d$ from 
\begin{equation}
P\propto d^{-3.7}
\label{powerlaw1}
\end{equation}
closer to the jet base to 
\begin{equation}
P\propto d^{-2.4} 
\label{powerlaw2}
\end{equation}
further downstream. The particular exponents of the power laws depend weakly on $\sigma_{\rm M}$.

We assume the equilibrium between jet and ambient medium pressure. In order to model a jet shape break position along the jet, 
we need to introduce the exerted pressure dependence on $r$, which we choose
in the power law form 
\begin{equation}
P_{\rm ext}=P_0\left(\frac{r}{r_0}\right)^{-b}. \label{Pscale}  
\end{equation}
{Such a pressure profile is consistent  with
Bondi flow \citep{QN00, Shch08, NF11} having $b\in(1.5;\;2.5)$ for different models, with the limiting value 2.5 for classical supersonic Bondi flow.}
This power law with $b\approx 2.0$ allows us to reproduce well both the parabolic jet form upstream the break and conical downstream. Using power laws \autoref{powerlaw1}, \autoref{powerlaw2}, and \autoref{Pscale},
we obtain for small distances $r$ (magnetically dominated 
regime)
\begin{equation}
d \propto r^{0.54} \,.
\label{l1}
\end{equation}
Accordingly, for large distances (saturation regime)
\begin{equation}
d \propto r^{0.83} \,.
\label{l2}
\end{equation}
As we see, qualitatively, the power indices are in good agreement with the observational
data. Thus, we are able to reproduce the jet boundary shape behaviour without introducing two different pressure profiles, as was done in \citep{Asada12}.
Having the reasonable pressure dependence on a distance, we reproduce both power laws in a jet shape. For example, for a central mass $M = 10^9 \, M_{\odot}$ and black hole spin $a_* = 0.5$ the light cylinder radius is $R_{\rm L}\approx 7\times 10^{-4}$~pc.
We also set the total magnetic flux in an outflow $\Psi_0=10^{32}\;{\rm G\,cm^{2}}$ \citep{ZCST14, Zdziarski-15, N17Fr}, which gives 
the value $B(r_{\rm g}) \approx 1400$~G. Thus, for these test parameters the jet width at the break, designated by a star in \autoref{plot3}, has typical values $0.2- 1.0$~pc in agreement with the observational results in \autoref{t:break}.

In dimensionless units the point of transition from one power law for pressure as a function of a jet width to the other is defined by one parameter only: the jet initial magnetization. In the equipartition regime the jet bulk Lorentz factor is $\gamma=\sigma_{\rm M}/2$. The observed kinematics in parsec-scale jets constrains the initial magnetization to a value $\lesssim 100$ \citep{MOJAVE_XIII}, while estimates for $\sigma_{\rm M}$ based on core-shift effect measurements provide the preferred value $\lesssim 20$ \citep{NBKZ15}. 
In dimensional units the jet width at the break depends also on BH mass and spin. The distance to a shape transition along the jet is determined by the total magnetic flux in a jet and the ambient medium pressure. We address the question of bounding these parameters in the next paper (in preparation).

\subsection{Magnetization}
\label{ss-magn}

In this subsection we check whether the break in a jet shape corresponds to the transition from the
magnetically-dominated into the equipartition regime. The jet magnetization is defined as the ratio
of Poynting flux 
\begin{equation}
{\bf S}=\frac{c}{4\pi}{\bf E}\times{\bf B}
\end{equation}
to particle kinetic energy flux
\begin{equation}
{\bf K}=\gamma m c^2 n {\bf u}_{\rm p},
\end{equation}
where $n$ is particle number density in the jet proper frame.
Using the standard expressions for ideal MHD velocities and electric and magnetic fields,
one obtains the following expression for the magnetization:
\begin{equation}
\sigma=\frac{|{\bf S}|}{|{\bf K}|}=\frac{\Omega_{\rm F}I}{2\pi c \gamma\mu\eta}.
\end{equation} 
Using the definitions of bulk Lorentz factor $\gamma$ and total current $I$,
we rewrite it as
\begin{equation}
\sigma=\Omega_{\rm F}\frac{L-\Omega_{\rm F}r_{\perp}^2E/c^2}{E-\Omega_{\rm F}L-{\cal M}^2E}.
\end{equation}

In order to check $\sigma$ along the jet, we calculate the maximal magnetization across the jet 
for each given distance $r$. The magnetization is always much less than the unity at the
jet axis and at the jet boundary. The first holds everywhere, since the Poynting flux behaves at the jet axis as
\begin{equation}
|{\bf S}|\propto I=\pi j r_{\perp}^2+o(r_{\perp}^2)
\end{equation} 
if the current density $j$ has no singular behavior at $r_{\perp}=0$. Thus, $\sigma\rightarrow 0$ at the axis.
The same holds for the boundary in a case of the full electric current closure. Due to specific choice of integrals $E(\Psi)$, $L(\Psi)$,
and $\Omega_{\rm F}(\Psi)$ (B17), the Poynting flux together with the magnetization reach their maximum values at $\Psi=\Psi_0/2$. It is at this magnetic field line the flow attains its highest Lorentz factor across the jet for the given distance from the central source. Thus,
we choose the maximal magnetization reaching approximately unity as a criteria of a flow attaining the
ideal MHD equipartition regime.

\begin{figure}
\centering
\includegraphics[width=\columnwidth, angle=0]{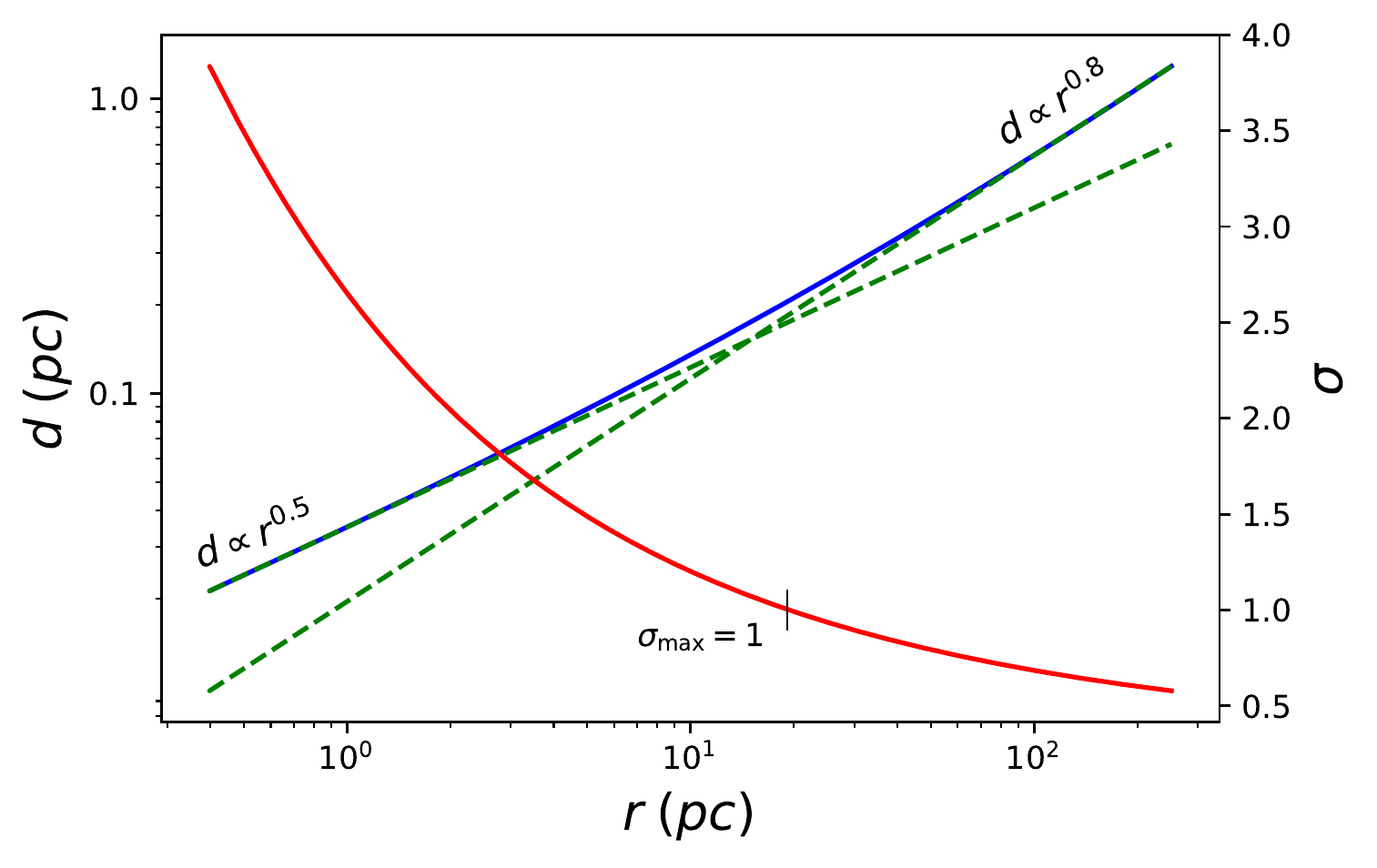}
\vskip -0.2cm
\caption{
An example of a jet boundary shape (blue solid line) for $\sigma_{\rm M}=50$ and $P_0=10^{-6}\;{\rm dyn/cm^2}$ at $r_0=10$~pc. The jet magnetization at a given distance from its base is plotted by a red solid line, with black vertical line marking $\sigma_{\rm max}=1$. 
The transition from one power law
to the other (green dashed lines) for the jet boundary roughly coincides with the point where the outflow transits from the
magnetically dominated to particle dominated (equipartition) regime.
}
\label{plotboundary}
\end{figure}

In \autoref{plotboundary} we present the maximal magnetization and the break in a jet form. We plot the modelled jet boundary shape for $\sigma_{\rm M}=50$, BH and jet parameters the same as in \autoref{ss:transition}.
The position of a jet shape break along a jet depends on an ambient pressure profile (\autoref{p0} and \autoref{Pscale}), and we use here, as an example, 
$P_0=10^{-6}\;{\rm dyn/cm^2}$ at $r_0=10$~pc.
We see that the break in jet shape 
occurs roughly at the distance from the BH, where the flow magnetization becomes equal to unity. For the higher initial magnetization it takes the larger transverse jet dimension in $R_{\rm L}$
to accelerate the flow up to equipartition, according to
\autoref{Gamma}.

\section{Discussion}
\label{s:discussion}
\subsection{Role of a Bondi sphere}
\label{ss:other_models}

\begin{figure}
\centering
\includegraphics[width=\columnwidth, angle=0]{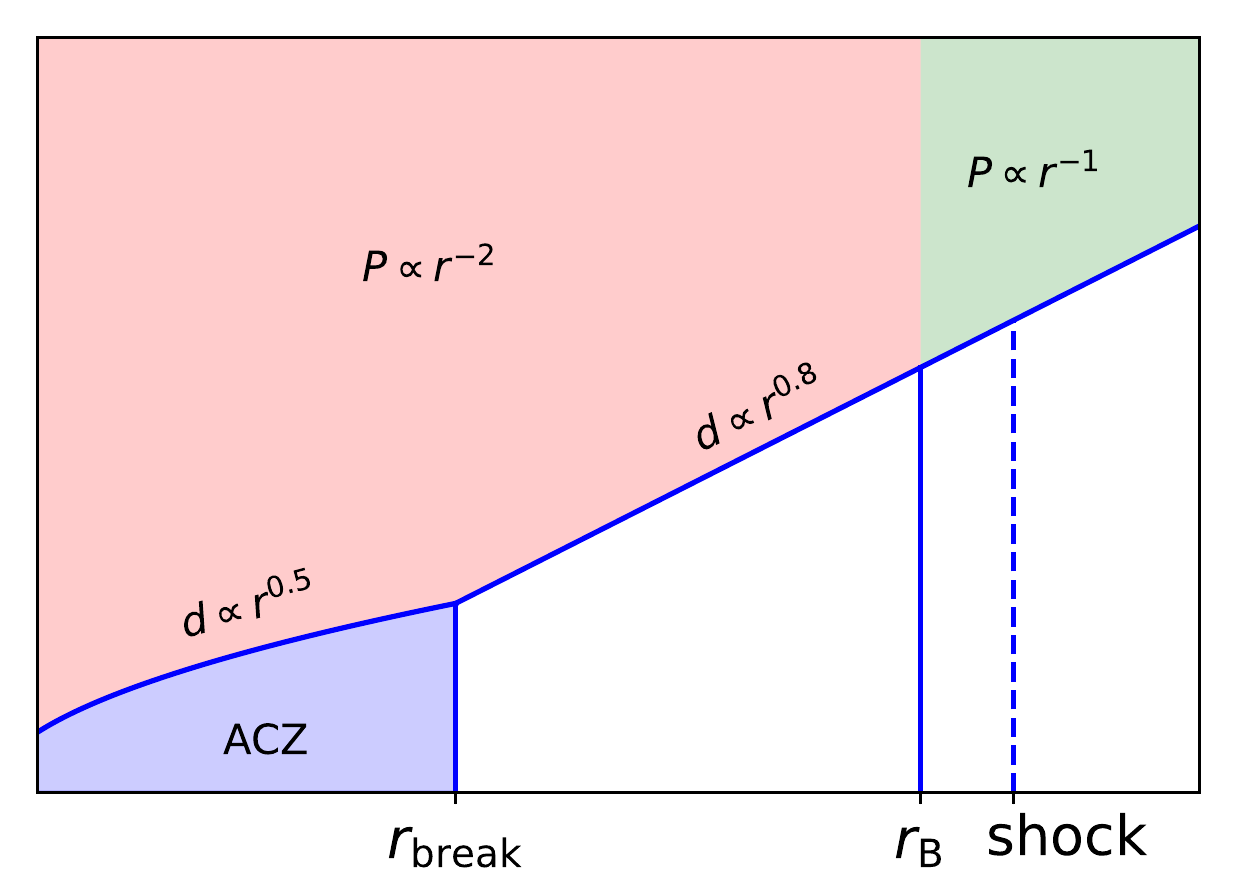}
\vskip -0.2cm
\caption{
A schematic jet boundary shape for an ambient pressure with different profiles, changing at the Bondi radius $r_{\rm B}$.
The jet accelerates while sustaining its boundary as a parabola (acceleration and collimation zone, ACZ). After reaching $\sigma=1$ at $r_{\rm break}$ the jet form becomes almost conical up to the Bondi radius. 
}
\label{crtn}
\end{figure}

In this paper we propose that the jet form change, observed in a dozen of nearby sources, may be explained by an internal flow transition from magnetically dominated to particle dominated regime with the smooth external pressure profile $P\propto r^{-2}$.
There are indications, however, that the ambient pressure may have different profiles at different scales. 
The measurements of particle number density in ISM by \citet{RF15} suggest
$\rho\propto r^{-1}$ from about 400~pc down to expected Bondi radius $r_{\rm B}\sim 100-250$~pc. The temperature profile on scales $100-1000$~pc is roughly constant. This means that just outside, or even inside, the Bondi radius, pressure profile is
$P\propto r^{-1}$, with no information on it inside a sphere $\sim 150$~pc.
The position of a sphere of influence is expected to be at a distance $10^5-10^6\;{r_\mathrm{g}}$ \citep{BMR-19}. The position of a transition point $r_\mathrm{break}$ from magnetically dominated to particle dominated regimes predicted by our model for reasonable parameters lay in general in the same interval or inside $r_\mathrm{B}$. 
For example, in the case of M87 we observe $r_\mathrm{break}\approx 40$~pc \citep{nokhrina2019} smaller than $r_\mathrm{B}$.
The same phenomenon has been noted by \citet{r:Nakahara18} for NGC~4261, where the structural transition lies well inside the expected sphere of influence. In \autoref{crtn} we present a cartoon for a jet shape with different ambient pressure profile.
Inside the Bondi sphere the jet is accelerating effectively up to the distance $r_\mathrm{break}$, with predicted parabolic boundary shape described by \autoref{l1}. This is the acceleration and collimation zone (ACZ) discussed by \citet{BMR-19}. For $r_\mathrm{break}<r<r_\mathrm{B}$ the jet assumes a close to conical form \autoref{l2}. Up to $r_\mathrm{B}$ the jet stays in equilibrium with the ambient pressure $P_\mathrm{jet}=P_\mathrm{ext}$. If for $r>r_\mathrm{B}$ the ambient pressure has a more shallow profile, the conical particle-dominated jet may become overpressured with a possible appearance of a standing shock. Thus, we predict the presence of a standing bright feature, associated with a shock, outside the Bondi sphere and downstream the break in jet shape. At this shock we may expect plasma heating, with the flow continuing a conical expansion \citep{BMR-19}.
The position of HST-1 in M87 jet in a close vicinity of expected $r_\mathrm{B}$ and downstream the $r_\mathrm{break}$ supports this picture.

\subsection{Additional observational evidence of the break point and predicted evolution of plasma acceleration}
\label{ss:compare-with-obs}

For each of the 10 sources with a jet geometry transition detected (\autoref{t:break}), we checked for slow pattern ($\beta_\textrm{app}<0.2c)$ jet features in \cite{MOJAVE_XVII}. We examined if their median locations with respect to the core are positionally associated, i.e., they match within the errors with the position of the derived jet shape break. We found that five sources have a quasi-stationary feature in the region where jet changes its shape, as expected (see discussion in \autoref{ss:other_models}). This is a factor of 1.5 larger compared to a ratio from the overall statistics of jet kinematics analysis performed at 15~GHz, which reveals a fraction of quasi-stationary jet features to be about 30\% \citep{MOJAVE_XVII}, applying the criterion $\beta_\textrm{app}<0.2c$. We underline that the MOJAVE kinematic analysis uses conservative criteria in cross-identifying components between epochs and selecting robust ones \citep{MOJAVE_XVII}. This means that the 50\% fraction of sources which show a standing feature in the break point region should be considered as a lower limit. This analysis is also conservative because of the requirement of the feature to be coincident with the detected break point. As discussed above, the shock may be located downstream the jet in the vicinity of $r_{\rm B}$, which position is usually not known.
We note that two sources included from other studies, 1H~0323+342 and M87, have the jet shape transition at distances larger than maximum angular scales probed by the MOJAVE 15~GHz observations.

\begin{figure}
\centering
\includegraphics[width=\columnwidth, angle=0]{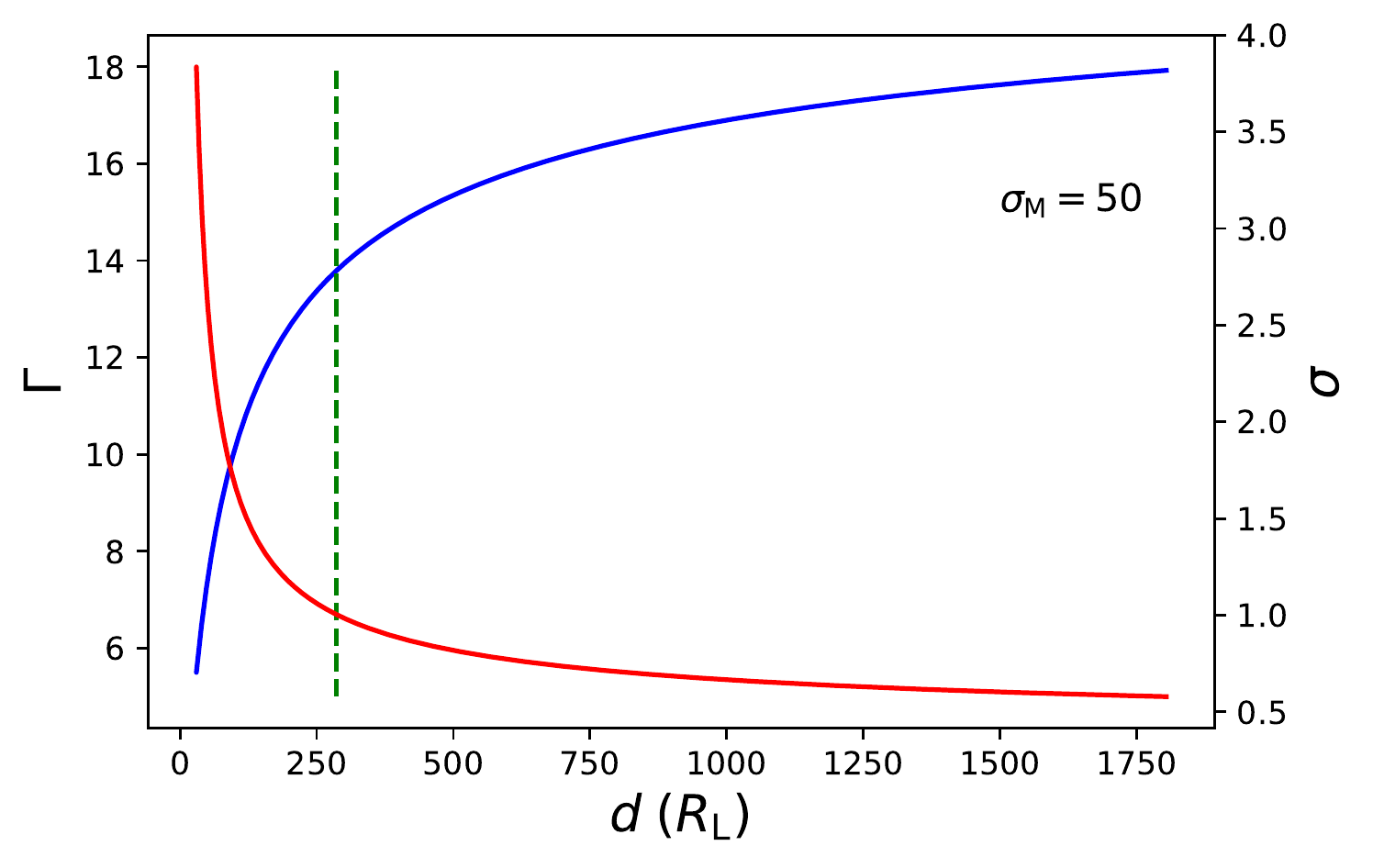}
\vskip -0.2cm
\caption{
An example of a Lorentz factor (blue solid line) growth against a deprojected distance along a jet in units of a light cylinder for $\sigma_{\rm M}=50$.
The red curve represents the maximum for a given jet cross-cut magnetization. The position of $\sigma=1$ is shown by a green dashed vertical line and it coincides with the break in a jet form. We predict that there must be an efficient plasma acceleration before the break, that becomes very slow after it.}
\label{f:gamma_predicted}
\end{figure}

We plot in \autoref{f:gamma_predicted} the maximum Lorentz factor of a bulk plasma motion along a jet, which we obtain within our semi-analytical model. The predicted pattern of a bulk Lorentz factor acceleration in magnetically dominated domain is $\gamma\propto r_{\perp}$, which provides for a parabolic jet $\gamma\propto r^{0.5}$. After the flow reaches equipartition, the acceleration continues slower than any power-law (logarithmically slow) \citep[e.g.,][]{Beskin06}. There is also a transitional zone between the two regimes. Thus, we would expect for the sources with the detected jet shape break and superluminal motion the following kinematics pattern: efficient Lorentz factor growth before the break point, and cessation of it in the conical region. This expected Lorentz factor behaviour was reported by \citet{Hada18}. The observed in radio band velocity map in M87 \citep{Mertens16} shows the acceleration saturation much earlier than the jet shape break. However, observations in the optical band \citep{BSM99} support the acceleration of plasma continuing further, with reported $\gamma=6$ at HST-1, situated downstream the jet shape break. This may point to non detection of fast components in radio.  

This prediction is consistent with observations by the MOJAVE program that acceleration is a common property of jet
features \citep[e.g.,][]{MOJAVE_XII,MOJAVE_XVII}, reflecting a tendency for increasing Lorentz factors near the
base of the jet, with decreasing or constant speeds being more common at projected distances $\gtrsim 10-20$ parsecs \citep{MOJAVE_XII}.
While decreasing speeds are not a prediction of this model for a change in jet shape, they could naturally occur if the reduction in positive acceleration is also accompanied by entrainment of external
material into the jet.

\section{Summary}
\label{s:summary}

\citet{MOJAVE_XIV} studied AGN jet shapes by measuring the power low index $k$ assuming a  $d\propto r^k$ dependence of the observed deconvolved jet width $d$ on the apparent distance from its core $r$. 
Most of the jets exhibited $k$ values in the range from 0.5 to 1.5.
As it was clearly demonstrated by \citet{MOJAVE_XIV}, high-quality, high-dynamic-range stacked images are needed for an analysis of this kind in order to trace the full jet channel. In view of a few recent exciting reports on jet shape transitions from parabolic to conical \citep[e.g.,][]{Asada12,Giroletti_etal08,tseng16, Hervet17,Hada18,r:Akiyama18,r:Nakahara18,r:Nakahara19}, we have performed a systematic search of such transition using MOJAVE 15~GHz stacked images, supplementing some of them with available single epoch 1.4~GHz VLBA images to trace larger scales.

Using an automated analysis approach, we have found 10 jets with such transition out of 367 analyzed:
0111$+$021,
0238$-$084,
0415$+$379,
0430$+$052,
0815$-$094,
1133$+$704,
1514$+$004,
1637$+$826,
1807$+$698,
2200$+$420.
Their redshifts lie in the range $z<0.07$ except for 0815$-$094, whose redshift is unknown. For the full analyzed sample the redshift values cover the range from 0.004 to 3.6 with the typical value being about 1.
This low-$z$ coincidence is unlikely to have occurred by chance. Taken together with an analysis of possible biases, we conclude that a genuine effect is present in the data for which VLBA reaches the linear resolution better than 1~pc. We would also predict that the BL~Lac object 0815$-$094 is a nearby AGN.

This finding leads to the following important conclusion.
A transition from parabolic to conical shape may be a general property of AGN jets. 
At the same time, we note that AGN observed at higher redshifts typically have higher luminosities and kinetic power, which can affect the collimation properties.
This conclusion has important implications for jet models, astrophysics and astrometry of AGN.
Measuring this phenomenon requires a search within nearby AGN which is the subject of our current followup study, or increasing the resolution by using Space VLBI \citep[e.g.,][]{RA_3C84} or high dynamic range high frequency VLBI imaging. 

The deprojected distance $r_\mathrm{break}$ from the nucleus to the break zone is found to be typically 10~pc. Even more interesting due to its relation to jet formation and acceleration models is this value measured in gravitational radius units. We find the range to be $r_\mathrm{break}\in(10^5,\,10^6)r_\mathrm{g}$ which corresponds to the typical Bondi radius.

We have developed the following model to explain the observed jet shape break. The accurate matching of a jet outflow with an ambient medium B17 predicts a change in jet shape from parabolic to conical if the ambient medium pressure is assumed to be governed by Bondi accretion. Within the model, a smaller external pressure is needed to support a jet than in earlier models. The transition of predicted jet shape from parabolic to conical occurs in the domain where the bulk plasma kinetic energy flux becomes equal to the Poynting energy flux, i.e., where the bulk flow acceleration reaches saturation \citep{Beskin06}. From studying the break properties we can estimate black hole spin and/or mass, jet total magnetic flux, and ambient medium properties as discussed by Nokhrina et al.~(in prep.).

The following two model predictions are supported observationally. The break point, where jets start to be plasma dominated energetically, might be a preferable domain for shocks.  We detect standing jet features in this region from MOJAVE analysis \citep{MOJAVE_XVII} in at least a half of the AGN targets. The plasma acceleration is predicted to decrease significantly at the transition region, which is consistent with MOJAVE acceleration results \citep{MOJAVE_XII,MOJAVE_XVII}.

Our finding also implies the following \citep[see also discussion in][]{algaba17}.
The well-known effect of the apparent shift of the core position with frequency due to synchrotron self-absorption does not follow the $r_\mathrm{core}\propto\nu^{-1}$ law all the way up to the true jet base, since a $-1$ power low index is expected only for a conical jet \citep{BK79,Lobanov98_coreshift}. Geometrical and physical estimates made on the basis of core shift measurements will need to take this into account while VLBI and VLBI-\textit{Gaia} astrometry applications will need to correct for it \citep{Porcas_cs2009} in cases where very high accuracy is required.

\section*{Acknowledgments}

We thank Dan Homan, Eduardo Ros, Luigi Foschini, Dave Meier as well as the anonymous referee for valuable comments and suggestions.
This study has been supported in part by the Russian Science Foundation: project 16-12-10481.
TS was supported by the Academy of Finland projects 274477, 284495, and 31249.
This research has made use of data from the MOJAVE database that is maintained by the MOJAVE team \citep{MOJAVE_XV}.
The MOJAVE program is supported under NASA-Fermi grant NNX15AU76G.
This study makes use of 43~GHz VLBA data from the VLBA-BU Blazar Monitoring Program, funded by NASA through the Fermi Guest Investigator Program. The VLBA is an instrument of the National Radio Astronomy Observatory. The National Radio Astronomy Observatory is a facility of the National Science Foundation operated by Associated Universities, Inc.
This research has made use of NASA's Astrophysics Data System.

\bibliographystyle{mnras}
\bibliography{yyk}

\onecolumn\pagestyle{empty}
\begin{landscape}
\setlength{\LTcapwidth}{\linewidth}
\setcounter{table}{0}
\begin{longtable}[p]{llcclrrrcrcr}
\caption{Properties for 12 sources with a detected jet shape break from this study (\autoref{f:geometry_transition}) as well as 
\citet[][1H~0323+342]{Hada18} and \citet[][M\,87]{Asada12}.
They are supplemented by the MOJAVE-1 sources for which redshift values, Doppler factor estimates, and robust jet shape fits (\autoref{t:mc_onefit_pars}) are available.
Columns are as follows:
(1) B1950 name;
(2) alias;
(3) optical class, where Q = quasar, B = BL Lac, G = radio galaxy, N = Narrow Line Seyfert~1 (NLSy1);
(4) redshift;
(5) literature reference for the data in column (4);
(6) maximum apparent radial speed from \citet{MOJAVE_XVII},
(7) variability Doppler factor from \citet{Hovatta_etal09};
(8) viewing angle;
(9) black hole mass estimated basing on assumption of virialized broad lines region (BLR) movement and correlation between the size of BLR and UV/optical luminosity;
(10) literature reference for the data in column (9);
(11) black hole mass estimated by a stellar velocity dispersion method and associated fundamental plane method (for 2200$+$420);
(12) literature reference for the data in column (11).
Names of the sources with the shape break are highlighted by the boldface font.
\label{t:sourprop}
}
\\
\hline\hline\noalign{\smallskip}
    Source &             Alias &Opt. &     $z$ &                  Reference & $\beta_\mathrm{app}$ & $\delta$ & $\theta$ &$M_{BH}$&                  Reference &$M_{BH}$&    Reference  \\
           &                   &  ID &         &                            &     (c) &          &    (deg) &($\log(M_{\sun})$)&                 &($\log(M_{\sun})$)&      \\
       (1) &               (2) & (3) &     (4) &                        (5) &     (6) &      (7) &      (8) &   (9) &                       (10) &   (11) &          (12) \\\hline\endfirsthead
\hline
       (1) &               (2) & (3) &     (4) &                        (5) &     (6) &      (7) &      (8) &   (9) &                       (10) &   (11) &          (12) \\\hline\endhead
0016$+$731 &    S5   0016$+$73 &   Q &   1.781 & \cite{1986AJ.....91..494L} &    8.22 &      7.8 &      7.4 &   9.18 &             \cite{Trrlb12} & \ldots &        \ldots \\
0059$+$581 &   TXS  0059$+$581 &   Q &   0.644 & \cite{2005ApJ...626...95S} &    6.84 & 21.9$^a$ &      1.5 & \ldots &                     \ldots & \ldots &        \ldots \\
0106$+$013 &    4C    $+$01.02 &   Q &   2.099 & \cite{1995AJ....109.1498H} &   10.91 &     18.2 &      2.8 & \ldots &                     \ldots & \ldots &        \ldots \\
\textbf{0111$+$021} &   UGC       00773 &   B &   0.047 &                \cite{WW76} &    0.17 &   \ldots &  5.0$^b$ & \ldots &                     \ldots & \ldots &        \ldots \\
0133$+$476 &    DA          55 &   Q &   0.859 & \cite{1986AJ.....91..494L} &   16.53 &     20.5 &      2.7 &   9.07 &             \cite{Trrlb12} &   \ldots &   \ldots \\
0215$+$015 &    OD         026 &   Q &   1.715 &  \cite{1988AA...192....1B} &   15.83 & 27.1$^a$ &      1.8 & \ldots &                     \ldots & \ldots &        \ldots \\
0234$+$285 &    4C    $+$28.07 &   Q &   1.206 & \cite{2012ApJ...748...49S} &   24.58 &     16.0 &      3.3 &   9.22 & \cite{2012ApJ...748...49S} & \ldots &        \ldots \\
\textbf{0238$-$084} &   NGC        1052 &   G &   0.005 & \cite{2005MNRAS.356.1440D} &    0.42 &  0.3$^a$ &     49.0 &   5.51 &             \cite{Trrlb12} &   8.19 &   \cite{WU02} \\
\textbf{0321$+$340} &    1H  0323$+$342 &   N &   0.061 & \cite{1996MNRAS.281..425M} &    9.02 &   \ldots &     6.3 &   7.30 &             \cite{Landt17} & \ldots &        \ldots \\
0336$-$019 &   CTA          26 &   Q &   0.852 & \cite{1978ApJS...36..317W} &   24.45 &     17.2 &      3.1 &   8.74 &             \cite{Trrlb12} &   \ldots &   \ldots \\
\textbf{0415$+$379} &    3C         111 &   G &   0.049 &\cite{EracleousHalpern2004} &    8.15 &  2.0$^a$ &     13.4 &   8.21 &             \cite{Trrlb12} & \ldots &        \ldots \\
0420$-$014 &   PKS   0420$-$01 &   Q &   0.916 & \cite{2009MNRAS.399..683J} &    2.36 &     19.7 &      0.7 &   8.98 &             \cite{Trrlb12} &   \ldots &   \ldots \\
\textbf{0430$+$052} &    3C         120 &   G &   0.033 & \cite{1988PASP..100.1423M} &    5.27 &  2.1$^a$ &     18.7 &   7.52 &             \cite{Trrlb12} &   8.13 &   \cite{WU02} \\
0458$-$020 &    S3   0458$-$02 &   Q &   2.286 & \cite{1974ApJ...190..509S} &    6.07 &     15.7 &      2.5 & \ldots &                     \ldots & \ldots &        \ldots \\
0528$+$134 &   PKS  0528$+$134 &   Q &   2.070 & \cite{1993ApJ...409..134H} &   10.91 &     30.9 &      1.2 &   9.03 &             \cite{Palma11} & \ldots &        \ldots \\
0605$-$085 &    OC      $-$010 &   Q &   0.870 & \cite{2012ApJ...748...49S} &   31.98 &      7.5 &      3.4 &   8.63 & \cite{2012ApJ...748...49S} & \ldots &        \ldots \\
0642$+$449 &    OH         471 &   Q &   3.396 & \cite{1994ApJ...436..678O} &    8.53 &     10.6 &      5.3 &   9.12 &             \cite{Trrlb12} & \ldots &        \ldots \\
0735$+$178 &    OI         158 &   B &   0.450 &  \cite{2012AA...547A...1N} &    5.04 &  4.5$^a$ &     12.7 & \ldots &                     \ldots & \ldots &        \ldots \\
0736$+$017 &    OI         061 &   Q &   0.189 & \cite{2009ApJS..184..398H} &   11.89 &      8.5 &      6.4 &   8.10 &             \cite{Trrlb12} &   \ldots &   \ldots \\
0754$+$100 &   PKS  0754$+$100 &   B &   0.266 &  \cite{2003AA...412..651C} &    1.06 &      5.5 &      4.0 & \ldots &                     \ldots & \ldots &        \ldots \\
0805$-$077 &   PKS   0805$-$07 &   Q &   1.837 & \cite{1988ApJ...327..561W} &   39.71 & 14.9$^a$ &      2.5 & \ldots &                     \ldots & \ldots &        \ldots \\
\textbf{0815$-$094} &   TXS  0815$-$094 &   B &  \ldots &                     \ldots &  \ldots &   \ldots &  5.0$^b$ & \ldots &                     \ldots & \ldots &        \ldots \\
0827$+$243 &    OJ         248 &   Q &   0.942 & \cite{2012ApJ...748...49S} &   19.81 &     13.0 &      4.0 &   8.77 &             \cite{Trrlb12} & \ldots &        \ldots \\
0851$+$202 &    OJ         287 &   B &   0.306 &  \cite{1989AAS...80..103S} &    6.59 &     16.8 &      2.3 & \ldots &                     \ldots & \ldots &        \ldots \\
0923$+$392 &    4C    $+$39.25 &   Q &   0.695 & \cite{2004AJ....128..502A} &    2.76 &      4.3 &     12.4 &   8.75 &             \cite{Trrlb12} &  \ldots &   \ldots \\
0945$+$408 &    4C    $+$40.24 &   Q &   1.249 & \cite{2004AJ....128..502A} &   20.20 &      6.3 &      5.2 &   8.66 &             \cite{Trrlb12} & \ldots &        \ldots \\
1055$+$018 &    4C    $+$01.28 &   Q &   0.888 & \cite{2012ApJ...748...49S} &    7.00 &     12.1 &      4.1 &   9.14 &             \cite{Trrlb12} & \ldots &        \ldots \\
1127$-$145 &   PKS   1127$-$14 &   Q &   1.184 & \cite{1986MNRAS.218..331W} &   18.93 & 21.9$^a$ &      2.6 &   9.03 &             \cite{Trrlb12} & \ldots &        \ldots \\
\textbf{1133$+$704} &  Mrk          180 &   B &   0.045 & \cite{1999PASP..111..438F} &  \ldots &   \ldots &  5.0$^b$ & \ldots &                     \ldots &   8.21 &   \cite{WU02} \\
1156$+$295 &    4C    $+$29.45 &   Q &   0.725 & \cite{2012ApJ...748...49S} &   24.59 &     28.2 &      2.0 &   8.35 &             \cite{Trrlb12} & \ldots &        \ldots \\
\textbf{1228$+$126} &     M          87 &   G &   0.004 & \cite{2000MNRAS.313..469S} &  \ldots &   \ldots &     14.0 & \ldots &                     \ldots &   9.82 &  \cite{Geb11} \\
1253$-$055 &    3C         279 &   Q &   0.536 & \cite{1996ApJS..104...37M} &   20.58 &     23.8 &      2.4 &   9.52 &             \cite{Trrlb12} &   \ldots &   \ldots \\
1308$+$326 &    OP         313 &   Q &   0.997 & \cite{2012ApJ...748...49S} &   21.30 &     15.3 &      3.5 &   8.76 &             \cite{Trrlb12} & \ldots &        \ldots \\
1413$+$135 &   PKS  1413$+$135 &   B &   0.247 & \cite{1992ApJ...400L..17S} &    1.20 &     12.1 &      0.9 & \ldots &                     \ldots & \ldots &        \ldots \\
1502$+$106 &    OR         103 &   Q &   1.839 & \cite{2008ApJS..175..297A} &   18.25 &     11.9 &      4.4 &   9.44 &             \cite{Trrlb12} & \ldots &        \ldots \\
1510$-$089 &   PKS   1510$-$08 &   Q &   0.360 & \cite{1990PASP..102.1235T} &   21.56 &     16.5 &      3.4 &   8.36 &             \cite{Trrlb12} & \ldots &        \ldots \\
\textbf{1514$+$004} &     PKS 1514$+$00 &   G &   0.052 &             \cite{CGRABS} &    0.45 &   \ldots & 15.0$^c$ & \ldots &                     \ldots &   \ldots &   \ldots \\
1538$+$149 &    4C    $+$14.60 &   B &   0.606 & \cite{2013ApJ...764..135S} &    8.75 &      4.3 &     10.5 & \ldots &                     \ldots & \ldots &        \ldots \\
1611$+$343 &    DA         406 &   Q &   1.400 & \cite{2012ApJ...748...49S} &   31.08 &     13.6 &      3.1 &   9.00 &             \cite{Trrlb12} &   \ldots &  \ldots \\
1633$+$382 &    4C    $+$38.41 &   Q &   1.813 & \cite{2012ApJ...748...49S} &   29.22 &     21.3 &      2.6 &   9.00 &             \cite{Trrlb12} & \ldots &        \ldots \\
1637$+$574 &    OS         562 &   Q &   0.751 & \cite{1996ApJS..104...37M} &   13.59 &     13.9 &      4.1 &   8.72 &             \cite{Trrlb12} &   \ldots &   \ldots \\
\textbf{1637$+$826} &   NGC        6251 &   G &   0.024 & \cite{2003AJ....126.2268W} &  \ldots &   \ldots &   18.0 &  \ldots  &                     \ldots &   8.78 & \cite{Frrrs99}\\
1641$+$399 &    3C         345 &   Q &   0.593 & \cite{1996ApJS..104...37M} &   19.28 &      7.7 &      5.1 &   9.00 &             \cite{Trrlb12} &   \ldots &   \ldots \\
1730$-$130 &  NRAO         530 &   Q &   0.902 & \cite{1984PASP...96..539J} &   13.73 &     10.6 &      5.2 & \ldots &                     \ldots & \ldots &        \ldots \\
1749$+$096 &    OT         081 &   B &   0.322 &  \cite{1988AA...191L..16S} &    7.91 &     11.9 &      4.5 & \ldots &                     \ldots & \ldots &        \ldots \\
1803$+$784 &    S5  1803$+$784 &   B &   0.680 & \cite{1996ApJS..107..541L} &    1.16 &     12.1 &      0.9 &   9.20 &             \cite{Trrlb12} & \ldots &        \ldots \\
\textbf{1807$+$698} &    3C         371 &   B &   0.051 &  \cite{1992AAS...96..389d} &    0.01 &      1.1 &      7.3 &   7.14 &             \cite{Trrlb12} &   8.51 &   \cite{WU02} \\
1828$+$487 &    3C         380 &   Q &   0.692 & \cite{1996ApJS..107..541L} &   13.03 &      5.6 &      7.4 &   8.59 &             \cite{Trrlb12} &   \ldots &  \ldots \\
1849$+$670 &    S4   1849$+$67 &   Q &   0.657 &  \cite{1993AAS..100..395S} &   21.43 &  8.1$^a$ &      4.7 &   8.83 &             \cite{Trrlb12} &   \ldots &   \ldots \\
1928$+$738 &    4C    $+$73.18 &   Q &   0.302 & \cite{1986AJ.....91..494L} &    7.55 &      1.9 &     14.2 &   8.62 &             \cite{Trrlb12} &   \ldots &   \ldots \\
2136$+$141 &    OX         161 &   Q &   2.427 & \cite{1974ApJ...190..271W} &    2.58 &      8.2 &      4.0 & \ldots &                     \ldots & \ldots &        \ldots \\
2145$+$067 &    4C    $+$06.69 &   Q &   0.999 & \cite{1991ApJ...382..433S} &    3.09 &     15.5 &      1.4 &   9.28 &             \cite{Trrlb12} & \ldots &        \ldots \\
\textbf{2200$+$420} &    BL         Lac &   B &   0.069 & \cite{1995ApJ...452L...5V} &    9.96 &      7.2 &      7.6 & \ldots &                     \ldots &   8.23 &   \cite{WU02} \\
2201$+$171 &   PKS  2201$+$171 &   Q &   1.076 & \cite{1977ApJ...215..427S} &   17.56 & 10.0$^a$ &      4.9 & \ldots &                     \ldots & \ldots &        \ldots \\
2201$+$315 &    4C    $+$31.63 &   Q &   0.295 & \cite{1996ApJS..104...37M} &    7.99 &      6.6 &      8.5 &   8.78 &             \cite{Trrlb12} &   \ldots &  \ldots \\
2223$-$052 &    3C         446 &   Q &   1.404 & \cite{1983MNRAS.205..793W} &   17.74 &     15.9 &      3.6 & \ldots &                     \ldots & \ldots &        \ldots \\
2227$-$088 &   PHL        5225 &   Q &   1.559 & \cite{2004AJ....128..502A} &    3.07 &     15.8 &      1.4 &   8.90 &             \cite{Trrlb12} & \ldots &        \ldots \\
2230$+$114 &   CTA         102 &   Q &   1.037 & \cite{1994ApJS...93..125F} &   17.73 &     15.5 &      3.7 & \ldots &                     \ldots & \ldots &        \ldots \\
2251$+$158 &    3C       454.3 &   Q &   0.859 & \cite{1991MNRAS.250..414J} &    3.46 &     32.9 &      0.4 &   9.07 &             \cite{Trrlb12} &  \ldots &  \ldots \\
\hline\noalign{\smallskip}
\multicolumn{12}{l}{$^a$ Doppler factor value is from \citet{LPA17}.}\\
\multicolumn{12}{l}{$^b$ Assumed $\theta$ value as typical for BL Lacs.}\\
\multicolumn{12}{l}{$^c$ Assumed $\theta$ value as typical for radio galaxies in the list which do not show a strong counter-jet.}\\
\end{longtable}

\bsp    
\label{lastpage}

\end{landscape}

\end{document}